\documentclass[aps,showpacs,twocolumn,superscriptaddress]{revtex4}
\usepackage{graphicx}
\begin{document}
\title{Wobbling excitations in  $^{156}$Dy and $^{162}$Yb}
\author{J. Kvasil}
\affiliation{Institute of Particle and Nuclear Physics, Charles
University, V.Hole\v sovi\v ck\'ach 2, CZ-18000 Praha 8, Czech Republic}
\author{R. G.~Nazmitdinov}
\affiliation{Departament de F{\'\i}sica,
Universitat de les Illes Balears, E-07122 Palma de Mallorca, Spain}
\affiliation{Bogoliubov Laboratory of Theoretical Physics,
Joint Institute for Nuclear Research, 141980 Dubna, Russia}

\date{\today}
\begin{abstract}
We study in the cranked Nilsson plus random phase approximation
low-lying quadrupole excitations of positive parity and negative signature
in $^{156}$Dy and $^{162}$Yb at high spins. Special attention is paid to
a consistent description of wobbling excitations and their identification
among excited states. A good agreement between available
experimental data and the results of calculations is obtained. 
We found that in $^{156}$Dy the lowest odd spin gamma-vibrational states  
transform to the wobbling excitations after the 
backbending, associated with the transition from axially-symmetric to nonaxial 
shapes. Similar results are predicted for $^{162}$Yb. 
The analysis of electromagnetic transitions, related to the wobbling excitations, 
determines uniquely the sign of the $\gamma$-deformation in  
$^{156}$Dy and $^{162}$Yb after the transition point.
\end{abstract}
\maketitle

\section{Introduction}
Deformation is an important ingredient of nuclear dynamics at low energies
\cite{BM75,Rag}. Regular rotational bands, identified  in spectroscopic data, are
most evident and prominent  manifestations of an anisotropy  of a spatial nuclear
density distribution.
While an axial deformation of a nuclear potential is
well established, there is a long standing debate on existence of the triaxial
deformation. A full understanding of this degree of freedom in nuclei may give
impact for other mesoscopic systems as well. In particular, the importance of
nonaxiality is discussed recently for metallic clusters \cite{nes} and atomic
condensates \cite{bir}. In nuclei, the nonaxial deformation involves
the orientation degree of freedom that could be exhibited as  wobbling
excitations \cite{BM75,JM79,mar2,OH01} and chiral rotation \cite{fra}.

Various models predict that nuclear ground states in the mass region
$150\leq A \leq 180$ are characterised by  axially deformed shapes.
With increase of an angular momentum, numerous calculations
demonstrate that  deformed nuclei undergo a shape transition from axially deformed shape
to nonaxial one (cf  Ref.\onlinecite{Rag}). Such a transition may be manifested as
a backbending, known since early seventies \cite{Jon}. 
There are different mechanisms responsible for the backbending, in particular,  
in the transitional nuclei with $N\sim 90$ (cf Refs.\onlinecite{fr90,Re,JK1}). 
For example, we found (see Ref.\onlinecite{JK05}) that in $^{156}$Dy such a transition 
can be explained as a result of vanishing of the $\gamma$-vibrational excitations 
in the rotating frame, while in $^{162}$Yb it is due to a rotational alignment of 
a neutron two-quasiparticle configuration. In the both cases, neither mechanisms 
do not provide, however, a non ambiguous proof of the onset of triaxial shapes.

Nowadays, according to a general notion, the analysis of specific low-lying excited 
states near the yrast line could shed light on existence of the nonaxiality. 
For nonaxial shapes one expects the appearance of low-lying vibrational states, that may be
associated with a classical wobbling motion. Such excitations (called wobbling excitations)
were suggested first by Bohr and Mottelson in rotating even-even nuclei \cite{BM75},
and analysed soon within a simplified microscopic approach in
Refs.\onlinecite{mic,D78} (see also Ref.\onlinecite{shi1} and references therein).
According to the microscopic approach \cite{JM79,mar2}, the wobbling excitations
are  vibrational states of the negative signature
built on the positive signature yrast (vacuum) state. Their characteristic feature
is collective E2 transitions with $\Delta I=\pm 1 \hbar$ between these and yrast states.
First experimental evidence of such states in odd Lu nuclei was reported
only recently \cite{OH01}.

Properties of the wobbling excitations at different angular
momenta may be studied within the asymmetric rotor model
(ARM) \cite{BM75} (see also Appendix). The extension of this model for odd nuclei 
has been used recently for the analysis \cite{ham} of experimental properties
of the second triaxial superdeformed band in $^{163}$Lu, that carries a few features
associated with the wobbling excitations. The classical dispersion equation for
the wobbling mode \cite{BM75}, with irrotational moments of inertia, was used to
describe a spectrum. The moments of inertia were fitted in order to reproduce 
the experimental data. As a result, a controversial $\gamma$-reversed moment of 
inertia  was introduced in Ref.\onlinecite{ham} to resolve difficulties with the 
data interpretation. To overcome this problem the alignment of odd quasiparticle 
has been taken into account later in Ref.\onlinecite{ham1}. It appears that the 
approach suggested in Refs.\onlinecite{ham,ham1} may explain some
tendencies, being, however, a crude approximation  to  a full physical  picture of
the observed phenomenon.

To explain the same data in $^{163}$Lu, a non self-consistent microscopic analysis,
based on the cranked Nilsson potential, was performed in Refs.\onlinecite{mat1, mat2}.
As a basic tool, the dispersion equation  for wobbling excitations, derived in
the time-dependent Hartree-Bogoliubov approach in Ref.\onlinecite{mar2}, has been used.
Based on the solution of the microscopic equation,
in Ref.\onlinecite{mat2} it was concluded that the pairing correlations do not
affect the wobbling excitations, and this  should be considered as a specific
feature related to this mode. On the other hand, it was also found that the
wobbling excitations are very sensitive to a single-particle alignment.
As it is known, the alignment decreases the pairing correlations. Therefore,
the question arises about the validity of this conclusion. 
Furthermore, the authors admitted that the kinematic moment of inertia $\Im_x$ 
was not described properly (see the discussion in Ref.\onlinecite{mat2}). 
In addition, the analysis \cite{mat1,mat2} is based on 
constant mean field parameters that are not related to the physical minima of 
the chosen potential at different rotational frequencies. In fact, the deformation 
parameters have been taken semi-quantitatively to reproduce roughly experimental data.
This may be not so crucial for the description of strongly deformed
nuclei, when the mean field parameters evolve slowly with a variation of
the rotational frequency. However, this approximation
becomes  questionable for the analysis of the rotational
bands in the backbending region, which are the main subject of our paper.
Evidently, mean field parameters can change drastically at the backbending
(see, for example, Ref.\onlinecite{JK05}).

We recall that wobbling excitations depend on  all three moments of inertia that
characterize the nonaxial shape. Therefore,  a self-consistent microscopic
description of the moments of inertia is a necessary requirement to the theory that
pretends to provide a solid analysis of experimental data.
Recently,  we developed a practical method based on the cranked Nilsson potential
with separable residual interactions  for the analysis of the low-lying excitations
near the yrast line.  In contrast to previous studies of low-lying excitations
at high spins (cf Refs.\cite{D78,shi1,mat1,mat2}), we paid a special attention
to the self-consistency between the mean field results and a description of
low-lying excitations in the random phase approximation (RPA).
Hereafter, we call our approach the CRPA. All the details of this approach are
thoroughly discussed in Ref.\onlinecite{JK05}. The analysis of M1-excitations \cite{JK2},
shape-phase transitions  and the behaviour of the
positive signature excitations at the backbending \cite{JK05}
confirmed the importance of the self-consistency for the description of moments of
inertia.  

In the present paper we will study  the low-lying negative signature
quadrupole vibrations and, in particular, the wobbling excitations in $^{156}$Dy 
and $^{162}$Yb that are subjected to the backbending at high spins. Note that a 
microscopic formulation of the wobbling excitations in the form of the macroscopic
rotor model was done by Marshalek in the time-dependent Hartree-Bogoliubov
approach for the spherically symmetric, pairing+quadrupole-quadrupole Hamiltonian 
in the principal axes (PA) frame \cite{mar2}. Following the same approach in the PA frame, 
Shimizu and Matsuzaki \cite{shi1} analysed the electromagnetic transitions 
and presented a few numerical examples, using the cranked modified harmonic oscillator 
(standard Nilsson) potential. In the standard Nillson potential, however, the 
inertial properties are not described correctly (see discussion in Ref.\onlinecite{JK05}). 
From point of view of the self-consistency, in order to study the vibrational 
excitations with this potential, it is appropriate to use the double-stretched 
quadrupole interaction \cite{SK}. We will provide a refined microscopic description 
of the wobbling excitations for doubly-stretched quadrupole interaction and their 
electromagnetic properties in a time-independent, uniformly rotating (UR) frame.
In  numerical analysis we will pay an attention  to the self-consistency 
between the mean field, vibrational excitations and their electromagnetic 
properties. We recall that in the UR frame one considers all negative signature 
excitations, including the wobbling one.
To identify the wobbling excitations in experimental data,  a few criteria are 
proposed in literature, such as: a large collectivity, zig-zag behavior of the  
B(E2) transition probability from a given band into the yrast one with 
$\Delta I=\pm 1$, when one of the transitions is almost dominant \cite{cas}.
We will present a microscopic procedure that provides a definite
answer how to identify the wobbling excitations. This procedure includes the
analysis of inertial properties, B(E2)- and B(M1)-transitions probabilities.

The paper is organized as follows:  in Section II we briefly review the main
details of our approach that is thoroughly discussed in Ref.\onlinecite{JK05}.
In Sec.III we study the lowest negative signature RPA excitations.
The main focus of this section is devoted to the definition of the specific
characteristics, associated with the wobbling mode.
This task will be done on the basis of the
comparison the CRPA approach  and the ARM.
The conclusions are finally drawn in Sec.\ IV.
To complete the analysis, we review properties of the wobbling mode in the 
ARM in Appendix.

\section{The Model}
\subsection{Basic properties of the mean field}
Our description is based on the Hamiltonian defined in the
UR frame
\begin{equation}
\label{3}
\hat H_{\Omega} \,=\, \hat H - \hbar \Omega \hat J_x \,=\,
\hat H_0 - \sum_{\tau=n,p} \lambda_{\tau} N_{\tau} - \hbar \Omega \hat J_x + V
\end{equation}
The unperturbed Hamiltonian $\hat H_0=\sum_{i} (h_{Nil}(i) + h_{add}(i))$ consists of
the Nilsson Hamiltonian
\begin{eqnarray}
\label{4}
h_{Nil} =  &&\frac{p^2}{2m} + \frac{1}{2} m (\omega_1^2 x_{1}^2 + \omega_2^2 x_{2}^2 +
\omega_3^2 x_{3}^2) \nonumber\\
&&-2 \kappa \hbar \omega_{00} {\bf l} \cdot {\bf s} -
\kappa \mu \hbar \omega_{00} ({\bf l}^{2} - \langle{\bf l}^{2}\rangle_N),
\end{eqnarray}
and the additional correction term \cite{nak}
\begin{eqnarray}
\label{4a}
h_{add} =
\Omega \,m\, \omega_{00} \kappa\Bigg{[}
&2&\left(r^2 s_x - x \vec r \cdot \vec s\right)\\
&+&\mu \left(2 r^2 - \frac{\hbar}{m\omega_{00}}
(N+\frac{3}{2})\right)\,l_x \Bigg{]}
\nonumber
\end{eqnarray}
The correction term restores  the local
Galilean invariance broken in the rotating coordinate system
and improves the description of the inertial properties in the
Nilsson model (see Ref.\onlinecite{JK05}). The chemical potentials 
$\lambda_\tau$ $(\tau=$n or p) are
determined so as to give correct average particle numbers $\langle \hat N_\tau \rangle$.
Hereafter, $\langle...\rangle$ means the averaging over the mean field vacuum (yrast)
state at a given rotational frequency $\Omega$. 

The two-body potential in Eq.(\ref{3}) includes the monopole pairing,
doubly-stretched quadrupole-quadrupole and monopole-monopole interaction and
spin-spin interaction. The Hamiltonian (\ref{3}) possesses
an inversion and signature symmetries.  
Using the generalized Bogoliubov transformation for quasiparticles
and  the variational principle (see details in Ref.\onlinecite{KN}), we obtain
the Hartree-Bogoliubov (HB) equations for the positive signature quasiparticle 
energies $\varepsilon _i $ (protons or neutrons).
The positive signature ($r=+1$) state is defined according to the Bogoliubov 
transformation
$\alpha_i^+=\sum_k({\cal U}_{ki} c_k^+ + {\cal V}_{\bar k i}c_{\bar k})$ and
$e^{i\pi j_x}\alpha_i^+ e^{-i\pi j_x}=-i\alpha_i^+$. Here,
$|k\rangle$ denotes a s.p. state of a Goodman spherical basis (see Ref.\onlinecite{JK}).
By diagonalizing the Hamiltonian at the rotational frequency $\Omega$, 
we obtain quasiparticle states with a good parity $\pi$ and  signature $r$.
It is enough to solve the HB equations for the positive
signature, since the negative signature eigenvalues and eigenvectors are
obtained from the positive ones
\begin {equation}
\left(-\varepsilon _i , {\cal U}_i, {\cal V}_i \right) \rightarrow \left( {\varepsilon} _{\bar{i}},
{\cal V}_{\bar{i}}, {\cal U}_{\bar{i}} \right)
\end {equation}
The index $\bar i$ denotes the negative signature $(r=-1)$ state
($e^{i\pi j_x}\alpha_{\bar i}^+ e^{-i\pi j_x}=i\alpha_{\bar i}^+$).
For a given value of the rotational frequency $\Omega$
the quasiparticle (HB) vacuum state is defined as
$\alpha_{i}|\rangle = \alpha_{\bar{i}}|\rangle = 0$. 

We solved a system nonlinear  HB equations for $^{156}$Dy and $^{162}$Yb
on the mesh of deformation parameters $\beta$ and $\gamma$ defined by dint 
of the oscillator frequencies in Eq.(\ref{4})  
\begin{equation}
\omega^{2}_{i} = \omega^{2}_{0}\left[ 1 - 2\beta \sqrt{\frac{5}{4\pi}}cos(\gamma
- \frac{2\pi}{3}i)\right], \,\, i = 1,2,3\, (or\, x,y,z)
\end{equation}
 The Nilsson-Strutinsky analysis of experimental data 
on high spins in $^{156}$Dy \cite{Kon} indicates that the positive parity yrast 
sequence undergoes a transition from the prolate (a collective rotation around the 
z-axes) towards an oblate (a non-collective rotation around the x-axes). In order to 
compare our results with available experimental data on excited states \cite{Kon}, 
in comparison with our previous work \cite{JK05}, we extended 
the range of the values for $\gamma$ from $\gamma=60^0$ (an oblate rotation around 
the y-axes) to $\gamma=-60^0$ (an oblate rotation around the x-axes). At each 
rotational frequency, in each mesh point we calculate self-consistently the total 
mean field energy $E_{HB}=\langle \hat H_{\Omega} \rangle$. In the vicinity of the backbending, 
the solution becomes highly unstable. In order to avoid unwanted singularities for certain
values of $\Omega$, we followed the phenomenological prescription \cite{wys}
\begin{equation}
\label{17}
\Delta_{\tau}(\Omega) \,=\,
\left \{
\begin{array}{l}
\Delta_{\tau}(0)\,[1-\frac{1}{2} (\frac{\Omega}{\Omega_c})^2 \,]
\qquad
\,\,\,\,\,\,\Omega < \Omega_c \\
\Delta_{\tau}(0)\,
\frac{1}{2} (\frac{\Omega_{c}}{\Omega})^2 \qquad \qquad \, \,
\,\,\,\,\,\,\,\,\Omega > \Omega_c, \\
\end{array},
\right.
\end{equation}
where $\Omega_c$ is  the critical rotational frequency
of the first band crossing.

\begin{figure}[ht]
\includegraphics[height=0.21\textheight,clip]{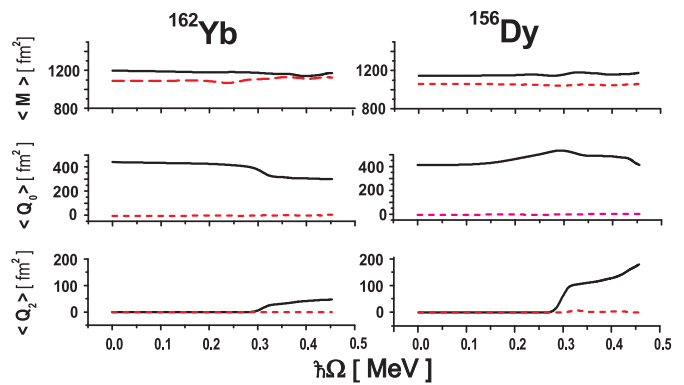}
\caption{
(Color online) The rotational behaviour of the calculated monopole and quadrupole moments.
The "double-stretched"  and standard values are connected by dashed and solid line,
respectively.}
\label{fig1}
\end{figure}

\begin{figure}[ht]
\includegraphics[height=0.22\textheight,clip]{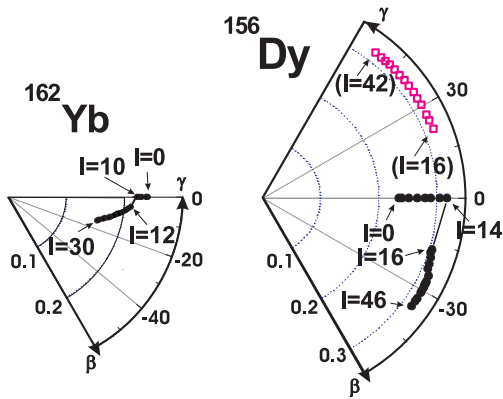}
\caption{
Equilibrium deformations in $\beta$-$\gamma$ plane
as a function of the angular momentum $I = \langle {\hat J}_x \rangle-1/2$
(in units of $\hbar$). The equilibrium deformations  for $^{156}$Dy provide the 
lower mean field energies in the region $-\pi/3<\gamma<0$ (filled circles)  
in comparison with those (open squares) obtained in Ref.\onlinecite{JK05}. 
The both branches of the equilibrium deformations are obtained enforcing the 
condition Eq.(\ref{new}). The maximal difference between the minimal HB energies 
at the positive and negative equilibrium $\gamma$-values does not exceed $\sim 1$MeV 
for $^{156}$Dy.}
\label{fig2}
\end{figure}

It is well known that for a deformed harmonic oscillator the quadrupole fields in
doubly-stretched coordinates fulfill the stability conditions
\begin{equation}
\langle\tilde Q_\mu\rangle = 0, \qquad \mu=0,1,2
\label{new}
\end{equation}
The tilde indicates that quadrupole fields are expressed in terms of 
doubly-stretched coordinates $\tilde{x}_i=(\omega_i/\omega_0)\,x_i$ and 
contain different combinations of the non-stretched quadrupole 
$Q_0\propto (2z^2-x^2-y^2)$, $Q_2\propto -\sqrt{3}(x^2-y^2)$ and monopole 
$M\propto r^2$ operators quantized along the axis z (cf \cite{SK}).
The condition Eq.(\ref{new}) holds, if the nuclear self-consistency
\begin{equation}
\omega_{1}^2 \langle x_{1}^2 \rangle = \omega_{2}^2 \langle x_{2}^2 \rangle=
\omega_{3}^2 \langle x_{3}^2\rangle
\label{scc}
\end{equation}
is satisfied in addition to the volume conserving  constraint.
In virtue of the condition (\ref{scc}) the doubly-stretched residual interaction
does not contribute to the mean field results in the Hartree procedure.
Enforcing the stability conditions Eq.(\ref{new}) in the Hatree-Bogoliubov approximation, 
we search the HB minimum for the Hamiltonian (\ref{3}) at a given rotational frequency.
While the mean field values of the quadrupole operators 
${\hat Q}_0$, ${\hat Q}_2$ are nonzero, the doubly-stretched 
quadrupole moments $ \langle{\tilde Q}_{0}\rangle$ and
$\langle{\tilde Q}_{2}\rangle$  are zeros (see Fig.\ref{fig1}) for  
equilibrium deformations (see Fig.\ref{fig2}). The deviation 
from the equilibrium values of the deformation parameters $\beta$ and $\gamma$  
results into a higher HB energy, indeed.

The results of our calculations conform to the results of the 
Nilsson+Strutinsky shell correction method (compare our Fig.\ref{fig2} with Fig.3c 
in Ref.\onlinecite{Kon}), although we obtain slightly different values for the 
equilibrium deformations. In the analysis of Ref.\onlinecite{Kon} the pairing 
correlations are omitted. In addition, in the Nilsson+Strutinsky shell correction 
method the rigid body moment of inertia simulates the inertial nuclear properties, 
which is different from the microscopic one, even at high spin region (see below). 
Moreover, the use of the Nilsson+Strutinsky results destroys 
the self-consistency between the mean-field calculations and the RPA analysis. 
Therefore, to keep a self-consistency between the mean field and the RPA as much 
as possible, we use the recipe described above. 

The triaxiality of the mean field sets in at
the critical rotational frequency $\Omega_c$ 
which triggers the backbending in the considered nuclei 
due to different mechanisms. As shown in Fig.\ref{fig2}, we obtain 
the critical rotational frequencies, at which
the first band crossing occurs, $\hbar \Omega_{c}\approx 0.25$MeV 
$(10\hbar\rightarrow 12\hbar)$ for $^{162}$Yb
and $\hbar \Omega_{c}\approx0.3$MeV 
$(14\hbar\rightarrow 16\hbar)$ for $^{156}$Dy. The contribution of the additional
term, Eq.(\ref{4a}), was crucial to achieve a good correspondence between the
calculated and experimental values of the crossing frequency in each nucleus
(see discussion in Ref.\onlinecite{JK05}).

In $^{156}$Dy  we obtain that the $\gamma$-vibrational excitations ($K=2$)
of the positive signature tends to zero in the rotating frame at the transition point,
in close agreement with experimental data. At the transition point
there are two indistinguishable  HB minima with different
shapes: axially symmetric and strongly nonaxial. 
It is interesting to note that this behaviour is symmetric with respect to the
sign of the $\gamma$-deformation, although the difference between the HB energy minima 
for $\gamma=\pm 20^0$ is about $\sim 0.8$MeV. The increase of the rotational
frequency changes the axial shape to the nonaxial one with a negative $\gamma$-deformation
($\gamma \sim -20^o$). The transition possesses all features of the shape-phase
transition of the first order. One may expect the appearance of the wobbling excitations near the
transition point, since after a shape transition there is a strong nonaxiality in
$^{156}$Dy. In contrast with $^{156}$Dy, in $^{162}$Yb
the axially symmetric configuration is replaced by the two-quasiparticle
one with a small negative $\gamma$-deformation. There, the backbending takes place due to
the rotational alignment of a neutron $i_{13/2}$ quasiparticle pair. The nonaxiality evolves quite
smoothly, exhibiting the main features of the shape-phase transition of the second order.
The question arises how is reliable our description or how self-consistently are
done our mean field calculations ?

\begin{figure}[ht]
\includegraphics[height=0.25\textheight,clip]{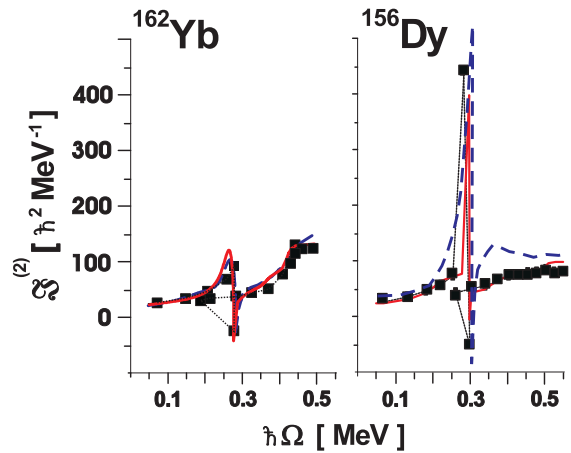}
\caption{
(Color online)
The rotational dependence of the dynamic
$\Im_{HB}^{(2)}=-d^2E_{HB}/d\Omega^2=d\langle{\hat J}_x\rangle/d\Omega$
(a solid line) and the Thouless-Valatin $\Im_{TV}$ moments of inertia
(a dash line). 
The experimental values $\Im^{(2)}=4/\Delta E_\gamma $ are denoted by
filled square connected by a thin line to guide eye. Here, $\Delta E_\gamma$ is the
difference between two consecutive
$\gamma$-transitions, and $E_\gamma$ is the $\gamma$-transition energy between
two neighboring states that differ on two units of the angular momentum}
\label{fig3}
\end{figure}
One of the conclusive tests of the self-consistency microscopic calculations is
the comparison of  the dynamic moment of inertia $\Im_{HB}^{(2)}$, calculated 
in the mean field approximation, and the Thouless-Valatin  moment of inertia 
$\Im_{TV}$, calculated in the RPA. They must coincide (see the results for exactly solvable model 
in  Ref.\onlinecite{n2} ), if one found a {\it self-consistent  mean field
minimum and spurious solutions are separated from the physical ones}. Our results 
(see Fig.\ref{fig3}) demonstrate a good self-consistency between the mean field 
and the CRPA calculations, indeed (see also the discussion in Ref.\onlinecite{JK05}).
We emphasize that the inclusion of the correction term, Eq.(\ref{4a}), is crucial
to achieve a good description of the inertial nuclear properties.

\subsection{Negative signature excitations}
To describe quantum oscillations around mean field solutions
the boson-like operators
$b^{+}_{k\bar{l}}=\alpha^{+}_{k}\alpha^{+}_{\bar{l}},
b^{+}_{kl}=\alpha^{+}_{k}\alpha^{+}_{l},
b^{+}_{\bar{k}\bar{l}}=\alpha^{+}_{\bar{k}}\alpha^{+}_{\bar{l}}$
are used.
The first equality introduces the positive signature boson,
while the other two determine the negative signature ones.
These two-quasiparticle operators are treated in the
quasi-boson approximation (QBA) as an elementary bosons, i.e.,
all commutators between them are approximated by their
expectation values with the uncorrelated HB vacuum \cite{RS}.
The corresponding commutation relations can be found in Ref.\onlinecite{KN}.
In this approximation the positive and negative signature boson spaces are
not mixed, since the corresponding operators commute and
$H_{\Omega}=H_\Omega(r=+1) + H_\Omega(r=-1)$. The analysis of the positive signature
term $H_\Omega(r=+1)$ is done in Ref.\onlinecite{JK05}.

In the UR frame, the negative signature RPA Hamiltonian has the form 
\begin{equation}
\hat H_{\Omega}[r =-1]=\frac{1}{2}\sum_{\mu}E_{\mu}b^{+}_{\mu}b_{\mu}-
\frac{\chi}{2}\sum_{\mu_{3}=1,2}\tilde{Q}^{(-)2}_{\mu_3}\, ,
\label{hns}
\end{equation}
where $E_{\mu} = \varepsilon_{i} + \varepsilon_{j}$
($E_{\bar i \bar j} = \varepsilon_{\bar i} + \varepsilon_{\bar j}$)
are two-quasiparticle energies.
Hereafter, we use the following definitions:
the index $\mu$ runs over $ij$, $\bar{i}\bar{j}$ and the
index $\mu_3$ is a projection on the quantization axis z.
The double stretched quadrupole operators
$\tilde{Q}_{1}^{(-)}=\xi {\hat Q}_1^{(-)}$ ($\xi=\omega_x \omega_z/\omega_0^2$),
$\tilde{Q}_{2}^{(-)}= \eta {\hat Q}_2^{(-)}$ ($\eta=\omega_x \omega_y/\omega_0^2$)
are defined by means of the quadrupole operators
${\hat Q}_m^{(r)}$ (m=0,1,2)
\begin{equation}
\hat Q_m^{(r)}=\frac{i^{2 + m +(r+3)/2}}{\sqrt{2(1+\delta _{m 0})}}
\biggl( \hat Q_{2m}+(-1)^{(r+3)/2}\hat Q_{2 -m}\biggr),
\label{xz}
\end{equation}
where $\hat Q_{\lambda m}={\hat r}^\lambda Y_{\lambda m}$.
We recall that the residual doubly-stretched interaction does not
distort the mean field deformations found self-consistently
for the Hamiltonian (\ref{3}). The violation of the self-consistency leads,
however, to unreliable description of the vibrational states.

The RPA Hamiltonian (\ref{hns}) contains only the isoscalar part of the quadrupole interaction,
since we would like to establish a connection between the microscopic approach
and the phenomenological ARM to see evidently similarities and differences.
Moreover, in the considered nuclei the main contribution of the isovector
quadrupole-quadrupole interaction is located in energy region around
$\sim 3$MeV and is responsible for M1 excitations \cite{JK2}.
One should keep in mind, however, that the isovector part of the quadrupole interaction
may be important for the analysis of wobbling excitations in odd-odd nuclei, when there is
a different orientation of the neutron and proton s.p. high-j orbitals and strong
$M1$-transitions are observed along yrast and/or yrare states.
The spin-spin interaction, $V_{\sigma \sigma} = - \frac{1}{2}
\sum_{T=0,1} \kappa_{\sigma} (T)\sum_{r=\pm} \sum_{\mu=0,1} (s_\mu [^{T}_{r}])^2$
(see the discussion in Refs.\onlinecite{JK2,JK05})
is not essential for low-lying quadrupole vibrations and is
omitted in our analysis. The pairing interaction does not contribute to
the boson Hamiltonian, $\hat H_{\Omega}[r =-1]$, since it is of the positive
signature. Notice, however, that the matrix elements of the operators depend on the
pairing interaction which affects the RPA solutions.

The linear boson part of the double stretched operators
has the following form
\begin{eqnarray}
\tilde{Q}_{1}^{(-)}&=&-\frac{1}{2}\sum_{\mu}\tilde{f_{1\mu}}(b^{+}_{\mu}+b_{\mu}),\quad
\tilde{f_{1\mu}}=\xi q_{1\mu}
\label{q1}\\
\tilde{Q}_{2}^{(-)}&=& -\frac{i}{2}
\sum_{\mu}\tilde{f_{2\mu}}(b^{+}_{\mu}-b_{\mu}),\quad
\tilde{f_{2\mu}}=\eta q_{2\mu}
\label{q2}
\end{eqnarray}
Here, $q_{1\mu}$, $q_{2\mu}$ are real matrix elements of the operators
${\hat Q}_1^{(-)}$, ${\hat Q}_2^{(-)}$, respectively
(the properties of matrix elements of the operators involved in the Hamiltonian
(\ref{hns}) can be found  in Ref.\onlinecite{JK}).

We solve the RPA equations of motion
\begin{eqnarray}
\label{10}
[\hat H_{\Omega}, \hat P_{\nu}] \,&=&\, i\,\omega_{\nu}\,\hat X_{\nu}, \qquad
[\hat H_{\Omega}, \hat X_{\nu}] \,=\, -i\,\omega_{\nu}\,\hat P_{\nu}, \nonumber\\
&&[\hat X_{\nu}, \hat P_{\nu'}]\,=\,i  \delta_{\nu\,\nu'},
\end{eqnarray}
where
\begin{equation}
\hat X_{\nu}=\sum_{\mu}X^{\nu}_{\mu}(b^{+}_{\mu}+b_{\mu}),\quad
\hat P_{\nu}=i\sum_{\mu}P^{\nu}_{\mu}(b^{+}_{\mu}-b_{\mu})
\end{equation}
are, respectively, the collective coordinates
and their conjugate momenta (hereafter, we use in all equations $\hbar=1$).
The RPA eigenfunction
\begin{eqnarray}
|\nu\rangle & = & \hat O_{\nu}^\dagger |RPA\rangle =
\frac{1}{\sqrt{2}}\,\Bigl(\,\hat X_{\nu} \,-\,i\hat{P}_{\nu} \,\Bigr)
|RPA\rangle\nonumber\\
&=&\sum_{\mu} (\psi^{(\nu)}_{\mu}b^{+}_{\mu}-\phi^{(\nu)}_{\mu}b_{\mu})|RPA\rangle
\label{phon}
\end{eqnarray}
defines the amplitudes $\psi^{(\nu)}_{\mu}$ and $\phi^{(\nu)}_{\mu}$
by means of the generalized coordinate and momentum amplitudes.
The ket vector $\mid RPA \rangle $ denotes the RPA vacuum (yrast state) at
the rotational frequency $\Omega$.

The solution of the above equations (\ref{10})
determines the generalized coordinate and momentum amplitudes
\begin{eqnarray}
X_\mu^\nu=\chi{\tilde R}^{\nu}_{1}
\frac{\omega_\nu {\tilde f}_{1\mu}}{E_\mu ^2-\omega_\nu^2}
+\chi {\tilde R}^{\nu}_{2}
\frac{E_\mu {\tilde f}_{2\mu}}{E_\mu ^2-\omega_\nu^2}\nonumber \\
 P_\mu^\nu=\chi{\tilde R}^{\nu}_{1}
\frac{E_\mu {\tilde f}_{1\mu}}{E_\mu^ 2-\omega_\nu^2}
+\chi{\tilde R}^{\nu}_{2}
\frac{\omega_\nu {\tilde f}_{2\mu}}{E_\mu ^2-\omega_\nu^2}
\label{XP}
\end{eqnarray}
with unknown coefficients
\begin{eqnarray}
{\tilde R}^{\nu}_{1} = \sum_{\mu}{\tilde f}_{1\mu}P^{\nu}_{\mu}\equiv
-\frac{1}{\sqrt{2}}\left[\hat O_\nu,\tilde Q_1^{(-)}\right]\nonumber\\
{\tilde R}^{(\nu)}_{2} = \sum_{\mu}{\tilde f}_{2\mu}X^{\nu}_{\mu}\equiv
\frac{i}{\sqrt{2}}\left[\hat O_\nu,\tilde Q_2^{(-)}\right]
\label{R12}
\end{eqnarray}
and eigenvalues $\omega_{\nu}$.
To find the eigenvalues $\omega_{\nu}$ one transforms
the system Eq.(\ref{XP}) to the form
\begin{eqnarray}
\nonumber
{\tilde R}^{\nu}_{1} \left[ D_{11} (\omega_{\nu}) - \frac{1}\chi\right] +
{\tilde R}^{\nu}_{2} D_{12}(\omega_{\nu}) &=& 0\label{seq1} \\
{\tilde R}^{\nu}_{1} D_{12}(\omega_{\nu}) + {\tilde R}^{\nu}_{2}\left[ D_{22} (\omega_{\nu})
- \frac{1}\chi\right] &=& 0
\end{eqnarray}
The condition
\begin{equation}
\label{det}
F(\omega_{\nu}) =  \det~({\bf D}- \frac{\bf 1}\chi)=0
\end{equation}
determines all negative signature RPA solutions. The matrix elements
$D_{km}(\omega_{\nu}) =\sum_\mu {\tilde f}_{k,\mu }
{\tilde f}_{m,\mu } C_\mu ^{km}/(E_\mu^2 -\omega_{\nu}^2)$
involve the coefficients $C_{\mu}^{km} = \omega_{\nu}$ for $k\neq m$ and
$E_{\mu }$ otherwise.
 Although the determinant has
a dimension $n=2$, one obtains a huge family of RPA solutions with different
degree of the collectivity. Among collective solutions there are
solutions that correspond to the shape fluctuations of the system.
Notice that  the direction of the angular momentum is fixed in the UR frame.

\section{The wobbling mode}
\subsection{Janssen-Mikhailov equation}
We recall that the doubly-stretched residual interaction restores in the RPA the
rotational symmetry broken at the mean field approximation. Therefore, 
for the cranking Hamiltonian the conservation laws  imply in the RPA order
\begin{equation}
[\,H_{\Omega}\,,\,\hat J_y\mp i\hat J_z \,]\, = \pm\Omega (\hat J_y\mp i\hat J_z)
\label{spur}
\end{equation}
This condition is equivalent to the condition of the existence of the negative 
signature solution $\omega_\nu=\Omega$ created by the operator 
(see Ref.\onlinecite{mar3})
\begin{equation}
\hat \Gamma^{\dagger}=\frac{\hat J_z + i \hat J_y}{\sqrt{2 \langle \hat J_x \rangle}},
\quad \hat \Gamma = (\hat \Gamma^{\dagger})^{\dagger}, \quad [\hat \Gamma, \hat \Gamma^{\dagger}]=1
\label{NG}
\end{equation}
This operator describes a collective rotational mode in the subspace of $\hat H_\Omega(-)$
\begin{equation}
[\hat H_\Omega(-),\hat \Gamma^{\dagger}]= \Omega \hat \Gamma^{\dagger},
\label{rot}
\end{equation}
arising from the symmetries broken by the external rotational field (the cranking 
term). 

\begin{figure}[ht]
\includegraphics[height = 0.3\textheight]{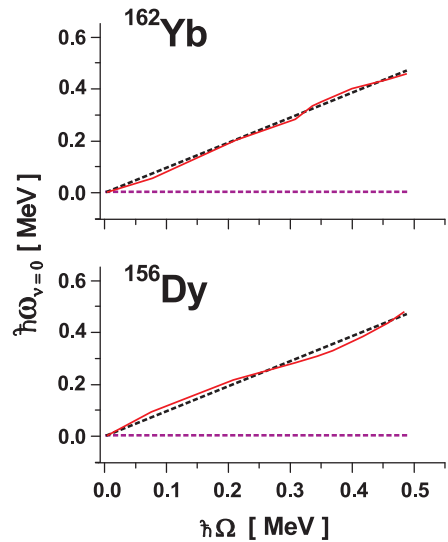}
\caption{
(Color online)
The evolution of the negative signature RPA solution $\omega_\nu=\Omega$ 
with (dashed line) and without (solid line) the additional term Eq.(\ref{4a})
as function of the rotational frequency, calculated at the equilibrium deformations
(see Fig.\ref{fig2}). The straight dashed line, parallel to the 
$\hbar\Omega$-axis, denotes the yrast line.}
\label{fig4}
\end{figure}

This is true for a pure harmonic oscillator model only \cite{n2}. However, our 
Hamiltonian (\ref{3}) contains the additional term $h_{add}$, Eq.(\ref{4a}).
Since the additional term $h_{add}$ has a term 
proportional to $\hat l_x$-operator, the conservation laws (\ref{spur}) are broken for 
the Hamiltonian (\ref{3}).  Nevertheless, these laws can be fulfilled in the RPA order 
by changing the strength constant in Eq.(\ref{det}) in order to obtain the solution 
$\omega_\nu=\Omega$ \cite{JK05}. To check this fact, we calculated the RPA secular 
equation (\ref{det}) for the mode $\omega_\nu=\Omega$, with and without the 
additional term $h_{add}$, for the same strength constant (see also Fig.10 in 
Ref.\onlinecite{JK05})). The results evidently demonstrate that the conservation 
laws (\ref{spur}), (\ref{rot}) are fulfilled with a good accuracy 
(see Fig.\ref{fig4}). In fact, due to a smallness of the term proportional to 
the operator $\hat l_x$ in the additional term, the violation is almost negligible. 

Based on this fact, we exploit the conservation laws (\ref{spur}) for the 
Hamiltonian (\ref{hns}),  
that yield the following equations
\begin{eqnarray}
\Omega  J^{z}_{\mu} + E_\mu J^{y}_{\mu}=
\chi\xi A {\tilde f}_{1\mu}
\label{ad1}\\
\Omega J^{y}_{\mu} + E_\mu J^{z}_{\mu}=
\chi\eta B {\tilde f}_{2\mu}
\label{ad2}
\end{eqnarray}
We will see below that these equations are fulfilled 
with a good numerical accuracy, since the redundant mode is separated from the 
other RPA solutions by our procedure for the strength constant. 
The parameters $A$ and $B$ 
\begin{eqnarray}
\xi A=\langle \,[{\tilde Q}_{1}^{(-)} \,,\,iJ_y\,]\rangle
=\sum_{\mu}{\tilde f}_{1\mu}J_{\mu}^y=
\xi\langle Q_{2}+\sqrt{3}Q_{0}\rangle
\label{sym1}\\
\eta B=\langle \,[{\tilde Q}_{2}^{(-)} \,,\,-iJ_z\,]\rangle
=\sum_{\mu}{\tilde f}_{2\mu}J_{\mu}^z=2
\eta\langle Q_{2}\rangle
\label{sym2}
\end{eqnarray}
 are obtained with the aid of the commutator
$[\hat J_x\pm i\hat J_y, \hat Q_{\lambda m}]=\sqrt{\lambda(\lambda+1)-m(m \pm 1)}\hat Q_{\lambda m \pm 1}$
(see Ref.\onlinecite{var}). The above relations 
between the matrix elements, Eqs.(\ref{ad1}),(\ref{ad2}), are
a key point for the analysis of the wobbling excitations at nonzero
$\gamma$-deformation, i.e., $\langle Q_2 \rangle\neq 0$. 
Moreover, in virtue of the definitions for
the phonon operator (Eqs.(\ref{phon}),(\ref{XP})) and
the operator $\hat \Gamma$ (Eq.(\ref{NG})), employing
Eqs.(\ref{ad1}),(\ref{ad2}), and Eq.(\ref{cond1}) below,  
one can show that
\begin{equation}
[\hat \Gamma,\hat O_\nu]=[\hat \Gamma,\hat O_\nu^\dagger]=0
\label{ort}
\end{equation}

Following the procedure described in Ref.\onlinecite{JM79}, 
with the aid of Eqs.(\ref{ad1}),(\ref{ad2}), one arrives to
the equation
\begin{equation}
(\omega_\nu^{2} - \Omega^{2}){\Delta}(\omega_{\nu})=0
\label{wob0}
\end{equation}
The determinant ${\Delta}(\omega_{\nu})$ corresponds to the system
of equations
\begin{eqnarray}
&\omega_\nu S_y r_1^\nu - \Omega {\Im}_{xy} r_2^\nu=0\nonumber\\
&\Omega {\Im}_{xz} r_1^\nu - \omega_\nu S_z r_2^\nu=0
\label{nzero}
\end{eqnarray}
for unknowns
\begin{equation}
r_1^\nu=\frac{\tilde R_1^\nu}{(\xi A)}, \quad r_2^\nu=\frac{\tilde R_2^\nu}{(\eta B)},
\label{x1}
\end{equation}
that does not contain anymore the solution $\omega_\nu=\Omega$.
We introduced above the following notations
\begin{eqnarray}
{\Im}_{xy}& = &\Im_x-\Im_y-\omega_\nu ^2S/{\Omega}\nonumber\\
{\Im}_{xz}& = &\Im_x-\Im_z-\omega_\nu ^2S/{\Omega}\label{not1}\\
S_{y,z}& = &{\Im}_{y,z}+\Omega S\qquad,\nonumber
\end{eqnarray}
where
\begin{equation}
S = \sum_\mu \frac{J^y_{\mu}J^z_{\mu}}{E^{2}_{\mu} -
\omega_\nu ^2}, \quad \Im_{y,z} = \sum_\mu
\frac{E_{\mu}{(J^{y,z}_\mu})^2}{E^{2}_{\mu} - \omega_\nu ^2}
\label{cmoi}
\end{equation}
and $\Im_x =  \langle \hat J_x \rangle/\Omega$ is the definition of the 
kinematic moment of inertia. 
From the system (\ref{nzero}), one obtains the relations
between unknowns $r_1^\nu$ and $r_2^\nu$
\begin{equation}
\frac{r_1^\nu}{r_2^\nu}=\frac{\Omega {\Im}_{xy}}{\omega_\nu S_y}=
\frac{\omega_\nu S_z}{\Omega {\Im}_{xz}},
\label{cond1}
\end{equation}
which are helpful for our analysis below.

The condition ${\Delta}(\omega_{\nu})=0$ leads to the Janssen-Mikhailov
equation \cite{JM79}
\begin{equation}
{\Delta}(\omega_{\nu})=\omega_\nu^2-\Omega^2\frac{[\Im_x - \Im_y -
\omega_\nu ^2 S/{\Omega}][\Im_x - \Im_z - \omega_ \nu^2 S/{\Omega}]}{[\Im_y +
\Omega S][\Im_z + \Omega S]}=0,
\label{wob1}
\end{equation}
that determines all vibrational modes of the negative signature
excluding the solution $\omega_\nu=\Omega$.
We stress that the solution of this equation alone is meaningless.
While Eq.(\ref{wob1}) does not depend on the strength constant,
the violation of the conditions (\ref{ad1}),(\ref{ad2}),
by arbitrary variation of the $\gamma$-deformation or pairing gap destroys
the link between the systems, Eq.(\ref{seq1}) and Eq.(\ref{nzero}).
As a result, the redundant mode can not be removed from Eq.(\ref{det}) 
and one is not able to obtain Eq.(\ref{wob1}).
Providing the wobbling solution, this equation has, however, a different form than the
Bohr-Mottelson classical equation. Below we present a simple
derivation of the microscopic analog of the latter one.

\subsection{Marshalek moments of inertia}
The relations Eq.(\ref{cond1}) are equivalent to the following ones
\begin{equation}
\frac{r_1^\nu S_y}{r_2^\nu S_z}=\frac{\Omega {\Im}_{xy}}{\omega_\nu S_z}=
\frac{\omega_\nu S_y}{\Omega {\Im}_{xz}}
\label{cond2}
\end{equation}
These relations can be associated
with the system of equations
\begin{eqnarray}
&\omega_\nu S_z a -  \Omega {\Im}_{xy} b = 0\nonumber\\
&\Omega {\Im}_{xz} a - \omega_\nu S_y b = 0
\label{nsys}
\end{eqnarray}
for unknowns $a=r_1^\nu S_y$ and $b=r_2^\nu S_z$. With the aid of definitions,
Eqs.(\ref{not1}), we rearrange this system to the following form

\begin{eqnarray}
&\omega_\nu (\Im_z+\omega_\nu S\frac{b}{a}) a -
\Omega (\Im_x-\Im_y-\omega_\nu S \frac{a}{b}) b=0\nonumber\\
&\Omega (\Im_x-\Im_z-\omega_\nu S \frac{b}{a}) a -
\omega_\nu (\Im_y+\omega_\nu S\frac{a}{b}) b=0
\label{mar}
\end{eqnarray}
In virtue of the relations, Eqs.(\ref{cond2}), between the unknowns $a$ and $b$,
we define the effective moments of inertia in the UR frame
\begin{eqnarray}
\Im_2^{\it eff}=\Im_y+\omega_\nu S\frac{a}{b}=\Im_y+\Omega S\frac{\Im_{xy}}{S_z}\nonumber\\
\Im_3^{\it eff}=\Im_z+\omega_\nu S\frac{b}{a}=\Im_z+\Omega S\frac{\Im_{xz}}{S_y}
\label{mmoi}
\end{eqnarray}
that depend on the RPA frequency. The definitions (\ref{mmoi}) are the same as 
those obtained by Marshalek in the time-dependent Hartree-Bogoliubov approach 
but in the PA frame \cite{mar2}. The determinant for nonzero solutions 
of the system Eq.(\ref{mar}) yields  the nonlinear equation similar to the 
classical expression for the wobbling mode \cite{BM75}
\begin{equation}
\omega_{\nu=w} = \Omega \sqrt{\frac{[\Im_{x} - \Im^{\it eff}_{2}]
[\Im_{x} - \Im^{\it eff}_{3}]}{\Im^{\it eff}_{2}\Im^{\it eff}_3}}\, ,
\label{wob2}
\end{equation}
with the microscopically defined moments of inertia.
Eq.(\ref{wob2}) was obtained first by Marshalek in the PA frame 
from the equations for the amplitudes of the angular frequency time oscillations. 
In the UR (time-independent) frame we obtain this equation with much less efforts 
as well.

\begin{figure}[ht]
\includegraphics[height=0.3\textheight,clip]{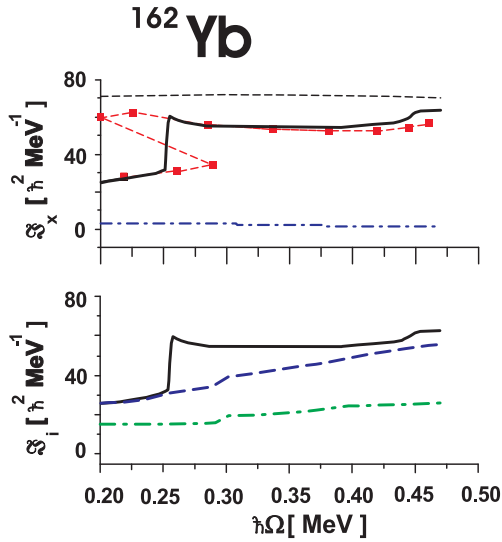}
\caption{
(Color online)
$^{162}$Yb. Top panel:
the kinematic moment of inertia
${\Im}_{x}= \langle{\hat J}_x\rangle/\Omega$
(solid line), a rigid body (dashed line)  and a hydrodinamical moments of inertia
(dash-dotted  line) are compared with the experimental values (filled squares).
Experimental values ${\Im}_{x}= I/\Omega$ are connected by dashed line to guide eyes 
($\hbar\Omega=E_{\gamma}/2$).
Bottom panel displays the rotational dependence of the kinematic moment of
inertia (solid line), Marshalek moments of inertia $\Im^{\it eff}_{2}$ (dashed line) 
and $\Im^{\it eff}_{3}$ (dash-dotted line) for the first RPA solution $\nu=1$ obtained from
Eq.(\ref{det}).}
\label{fig5}
\end{figure}

It is evident that for the rotation around the axis x
the wobbling excitations with different collectivity could be found from
Eq.(\ref{wob2}), if the condition
\begin{equation}
\Im_{x}>\Im^{\it eff}_{2},\Im^{\it eff}_{3} \quad (or
\quad \Im_{x}<\Im^{\it eff}_{2},\Im^{\it eff}_{3})
\label{cond3}
\end{equation}
is fulfilled. One may expect that for the RPA solutions, different from the
wobbling mode, this condition should not hold.
Notice  that Eq.(\ref{det}) contains the solutions to
Eq.(\ref{wob2}) but not {\it vise versa}, since the above constraint, 
Eq.(\ref{cond3}), valid for the later case, is not required for the former one.

\begin{figure}[bt]
\includegraphics[height=0.3\textheight,clip]{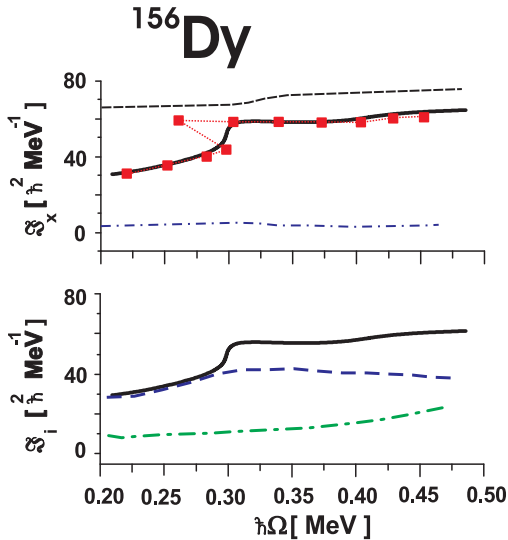}
\caption{
(Color online)
$^{156}$Dy. Similar to Fig.\ref{fig5}.}
\label{fig6}
\end{figure}
We obtain quite a remarkable correspondence between the experimental and calculated
values for the kinematic moment of inertia for the both nuclei (see top panels in 
Figs.\ref{fig5},\ref{fig6}). We trace also the evolution of the irrotational fluid 
moment of inertia (see Ref.\onlinecite{RS})
\begin{equation}
\Im^{(irr)}_{i}= \frac{3}{2\pi}mAR^{2}\beta^{2}\sin ^{2} \left( \gamma -
\frac{2\pi}{3}i \right)
\end{equation}
and a rigid body moment of inertia $(i=1,2,3)$
\begin{equation}
\Im^{(rig)}_{i} = \frac{2}{5} m AR^{2} \left( 1- \sqrt{\frac{5}{4\pi}} \beta
\cos (\gamma - \frac{2\pi}{3}i) \right)
\end{equation}
as a function of the equilibrium deformations (see Fig.\ref{fig2}).
The irrotational fluid moment of inertia $\Im^{(irr)}_1$
does not reproduce neither the rotational dependence nor the absolute values of the
experimental one as a function of the rotational frequency. 
The rigid body values provide the asymptotic limit of fast rotation without pairing.
Evidently, the difference between the rigid body and the calculated kinematic 
moments of inertia in the both nuclei decreases with the increase of the rotational
frequency, although it remains visible at high spins.
At very fast rotation $\hbar\Omega>0.45$MeV the pairing correlations are
reduced due to multiple alignments, and, therefore, the difference is moderated.

\begin{figure}[ht]
\includegraphics[height=0.25\textheight,clip]{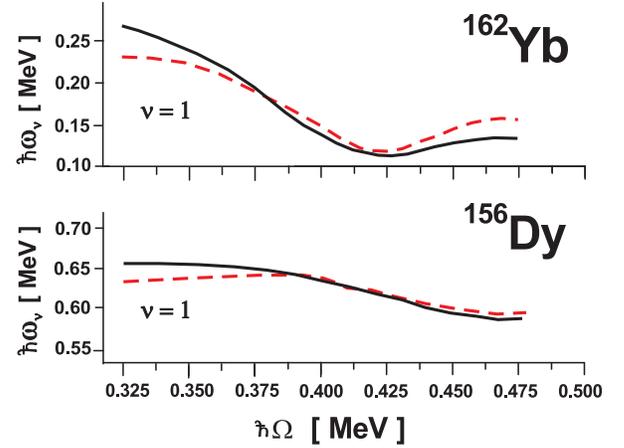}
\caption{(Color online)
The rotational dependence of the RPA solutions, obtained
with the aid of Eq.(\ref{det}) (solid line) and
Eq.(\ref{wob2}) (dashed line).
The RPA solutions, obtained with the aid of Eq.(\ref{det}),
fulfil the condition Eq.(\ref{cond3}). }
\label{fig7}
\end{figure}

The calculated values of the Marshalek moments of inertia, Eq.(\ref{mmoi}), signal
(see the bottom panels, Figs.\ref{fig5},\ref{fig6}) that one should
expect the appearance of the wobbling mode after a shape-phase transition in
$^{162}$Yb and $^{156}$Dy for the first RPA solution found with the aid Eq.(\ref{det}).
As was stressed above, the separation of the redundant mode is an essential point
for the RPA wobbling theory, that secures a reliable analysis of the RPA modes.
Additionally, to be assure in the self-consistency of our RPA 
calculations, we compare the solutions, that may be associated with
wobbling excitations, from different RPA equations (\ref{det}) and (\ref{wob2}).
We recall that Eq.(\ref{det}) depends on the strength constant $\chi$ and contains 
different RPA solutions including the redundant mode, while this dependence
is removed from Eq.(\ref{wob2}). Evidently, if the redundant mode would not be removed 
from Eq.(\ref{det}) by our choise of the strength constants, the conditions 
Eqs.(\ref{ad1}),(\ref{ad2}) should be broken. As a result, the consistency between
Eq.(\ref{det}) and Eq.(\ref{wob2}) would be broken as well and these equations 
would provide different solutions. A nice agreement
between the roots of Eq.(\ref{det}) and Eq.(\ref{wob2}) (see Fig.\ref{fig7})
confirms the vitality and the validity of our approach.

Does the constraint (\ref{cond3}) guarantee the existence of a collective
wobbling mode solution ? Below we formulate criteria 
how to identify the collective wobbling excitations.

\subsection{Criteria for wobbling excitations}
By means of  Eqs.(\ref{mar}),(\ref{mmoi}), we can define the unknown
variables $r_1^{\nu=w}=a/S_y$,  $r_2^{\nu=w}=b/S_z$
(which determine the wobbling mode) such that
\begin{eqnarray}
\frac{r_1^{w}}{r_2^{w}}&=&\frac{S_z}{S_y}\frac{a}{b}=
\frac{S_z}{S_y}\frac{\Omega(\Im_x-\Im^{\it eff}_2)}{\omega_w \Im^{\it eff}_3}=\nonumber\\
&=&\frac{S_z}{S_y}\frac{\omega_w \Im^{\it eff}_2}{\Omega(\Im_x-\Im^{\it eff}_3)}=
\sqrt{\frac{W_2}{W_3}}\frac{S_z}{S_y}\frac{\Im^{\it eff}_2}{\Im^{\it eff}_3}
\label{cond4}
\end{eqnarray}
Here, we used the fact that the dispersion equation for the
wobbling mode, Eq.(\ref{wob2}), can be expressed as
$\omega_w=\langle {\hat J}_x \rangle \sqrt{W_2W_3}$,
where
\begin{equation}
W_2=(1/\Im_2^{\it eff}-1/\Im_x), \quad W_3=(1/\Im_3^{\it eff}-1/\Im_x)
\label{maw}
\end{equation}
With the aid of Eqs.(\ref{mmoi}) and  the definition, Eqs.(\ref{not1}),
it is easy to show that the ratio $S_z\Im^{\it eff}_2/S_y\Im^{\it eff}_3\equiv1$. 
It results in the following relation
\begin{equation}
\frac{r_1^{w}}{r_2^{w}}=\sqrt{\frac{W_2}{W_3}}
\label{cond5}
\end{equation}
Our next task is to find a formal expressions for the unknowns $r_{1,2}^w$
in terms of the functions $W_{2,3}$ that depend on the microscopic moments of inertia.

By means of the inverse transformation
\begin{equation}
b_\mu^\dagger=\sqrt{2}\sum_\nu X_\mu^\nu (\hat O_\nu^\dagger-\hat O_\nu) +
P_\mu^\nu (\hat O_\nu^\dagger + \hat O_\nu)
\end{equation}
we can express the operators $\tilde Q_{1,2}^{(-)}$
(see Eqs.(\ref{q1}),(\ref{q2}), (\ref{R12}))
via phonon operators
\begin{eqnarray}
&\tilde Q_1^{(-)} = -\sqrt{2}\sum_{\nu}{\tilde R}_1^\nu (\hat O_\nu^\dagger + \hat O_\nu)\\
&\tilde Q_2^{(-)} = -i\sqrt{2}\sum_{\nu}{\tilde R}_2^\nu (\hat O_\nu^\dagger - \hat O_\nu)
\end{eqnarray}

Let us use the fact that the components of the quadrupole tensor commute, i.e.,
the condition
\begin{eqnarray}
&\left[\tilde Q_1^{(-)}, \tilde Q_2^{(-)}\right] =
4 i \sum_{\nu={\it all}} {\tilde R}_1^\nu {\tilde R}_2^\nu\nonumber\\
&= 2\sum_{\nu={\it all}} \left[{\hat D}_\nu, \tilde Q_1^{(-)}\right] \left[{\hat D}_\nu, \tilde Q_2^{(-)}\right]
=0
\label{cond6}
\end{eqnarray}
Here, we use the notation ${\hat D}_{\nu=\Omega}\equiv{\hat \Gamma}$ and 
${\hat D}_{\nu \neq\Omega}\equiv {\hat O}_\nu$
for other vibrational modes.
Taking into account the definitions Eqs.(\ref{R12}),(\ref{x1})
(see also Eqs.(\ref{NG}),(\ref{sym1}),(\ref{sym2})),
we obtain the exact definitions for the unknowns $r_{1,2}^\Omega$ associated with
the redundant mode:
\begin{eqnarray}
&&r_1^\Omega=\frac{{\tilde R}_1^{\nu=\Omega}}{\xi A} =
-\frac{1}{\xi A\sqrt{2}}\left[{\hat\Gamma},\tilde Q_1^{(-)}\right]=
-\frac{1}{2\sqrt{\langle \hat J_x \rangle}}, \nonumber\\
&&r_2^\Omega = \frac{{\tilde R}_2^{\nu=\Omega}}{\eta B} =
\frac{i}{\eta B\sqrt{2}}\left[{\hat\Gamma}, \tilde Q_2^{(-)}\right] =
\frac{1}{2\sqrt{\langle \hat J_x \rangle}}
\label{omega}
\end{eqnarray}
 This result yields the following expression for the sum, Eq.(\ref{cond6}),
\begin{equation}
\sum_{\nu\neq w, \Omega}r_1^\nu r_2^\nu + r_1^wr_2^w = -r_1^\Omega r_2^\Omega =
\frac{1}{4\langle \hat J_x \rangle}
\label{sum}
\end{equation}
Suppose, that in the l.h.s of Eq.(\ref{sum}) the sum, defined by all physical 
solutions excluding the wobbling one,
is zero due to a mutual cancellation of different terms.
Below, we will see that it is, indeed, the case.
As a result, we obtain the equation for the unknowns $r_{1,2}^w$.
Resolving this equation by dint of Eq.(\ref{cond5}), we obtain 
\begin{equation}
r_1^w=\frac{1}{2\sqrt{\langle \hat J_x\rangle}} \left(\frac{W_2}{W_3}\right)^{1/4}, \quad
r_2^w=\frac{1}{2\sqrt{\langle \hat J_x\rangle}} \left(\frac{W_3}{W_2}\right)^{1/4}
\label{wob4}
\end{equation}
These expressions are similar to those of the wobbling mode in the
Bohr-Mottelson model (see Appendix, Eqs.(\ref{clr})), although the quantities
$W_{2,3}$ are determined by the Marshalek moments of inertia and
$\langle \hat J_x \rangle \simeq I+1/2$ (the factor $1/2$ is arising due to the 
RPA contribution of the redundant mode, see discussion in Ref.\onlinecite{al}).
Note that similar expressions were obtained in Ref.\onlinecite{shi1} in the 
PA frame (with the quantization condition $\langle \hat J_x \rangle = I$ and some 
additional phases).

To identify the wobbling mode among the solutions of Eq.(\ref{det})
it is convenient to transform Eq.(\ref{cond6})  to the form
\begin{equation}
\sum_{\nu=all} c_\nu = 0, \quad
c_\nu=4\langle \hat J_x\rangle \frac{{\tilde R}_1^{\nu}}{\xi A}
\frac{{\tilde R}_2^{\nu}}{\eta B}
\label{coef}
\end{equation}
From Eqs.(\ref{omega}), (\ref{wob4}) it follows that
\begin{equation}
c_{\nu=\Omega} \equiv -1, \quad c_{\nu=w} \equiv 1
\label{crit}
\end{equation}
Thus, solving {\it only} the system of the RPA equations for the quadrupole
operators, Eq.(\ref{det}), the condition Eq.(\ref{crit}) enables us to
identify the redundant and the wobbling modes.

\subsection{Analysis of experimental data}
The experimental level sequences for all observed up-to-date 
rotational bands in $^{162}Yb$ and $^{156}$Dy are
taken from Ref.\onlinecite{nndc}, which is regularly updated.
The rotational bands are numerated 
as rotational bands B1, B2... in accordance with \cite{nndc}.
All rotational states are classified by the
quantum number $\alpha$ which is equivalent to our signature $r$.
The positive signature states ($r=+1$) correspond to $\alpha=0$, since
the quantum number $\alpha$ leads to
selection rules for the total angular momentum
$I=\alpha + 2n$, $n=0,\pm 1, \pm 2 \ldots $ (cf Ref.\onlinecite{fra}).
In particular, in even-even nuclei the yrast band
characterized by the positive signature quantum number
$r=+1 \,(\alpha = 0)$ consists of even spins only.
The negative signature states ($r=-1$) correspond to $\alpha=1$ and are 
associated with odd spin states in even-even nuclei. All considered bands are 
of the positive parity $\pi=+$.

\begin{figure}[ht]
\includegraphics[height=0.43\textheight,clip]{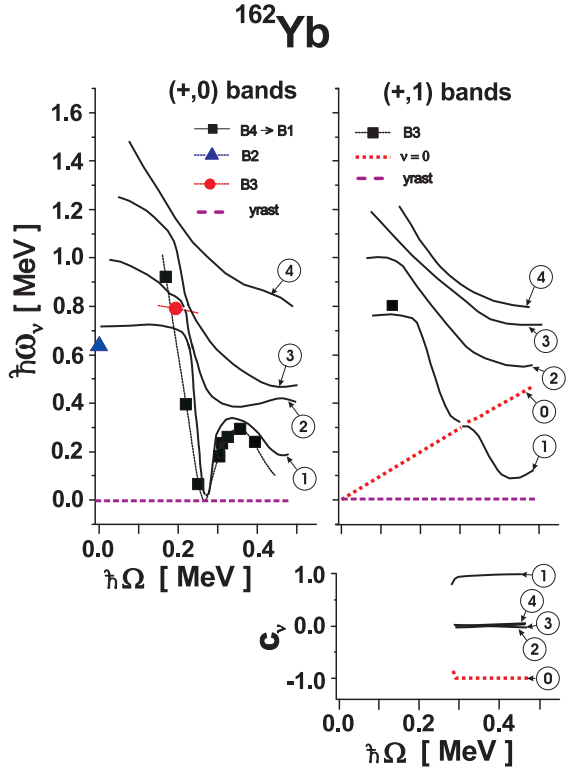}
\caption{(Color online)$^{162}$Yb.
Top left panel: the rotational dependence of the positive 
signature RPA solutions with even spins $(\pi=+,\alpha=0)$. Number in a circle denotes 
the RPA solution number : $1$ is the first $\nu=1$  RPA solution  etc.
Different symbols display the experimental data 
associated with B1,B2...bands (the band labels are taken in 
accordance with the definitions given in Ref.\onlinecite{nndc}).
Top right panel: the rotational dependence of the negative signature 
RPA solutions  with odd spins $(\pi=+,\alpha=1)$. The redundant mode 
$\omega_\nu=\Omega$ is denoted as "0" and is displayed by the dotted
line. Bottom panel: the rotational dependence of the coefficients
$c_\nu \sim \tilde R_{1}^\nu \tilde R_2^\nu$ (see Eq.(\ref{coef})) that are determined
by the solutions of Eq.(\ref{det}). }
\label{fig8}
\end{figure}

\begin{figure}[ht]
\includegraphics[height=0.43\textheight,clip]{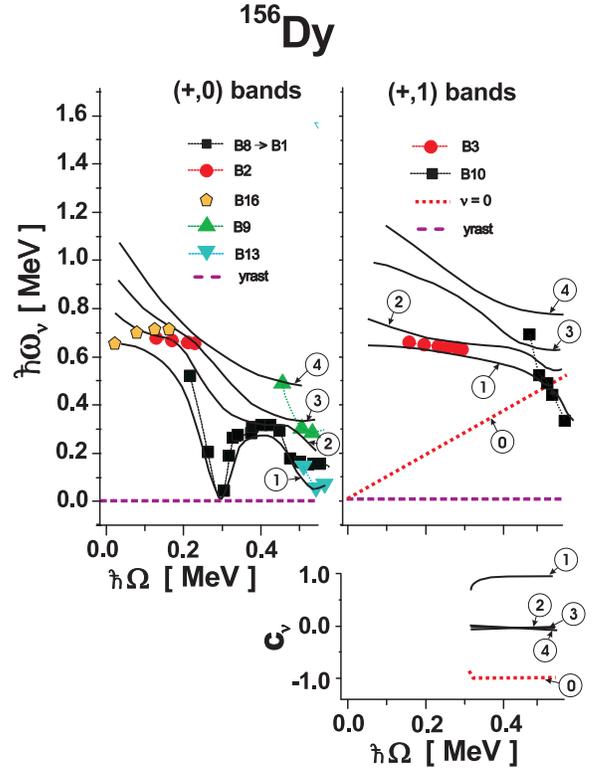}
\caption{(Color online)$^{156}$Dy. The same as for Fig.\ref{fig8}. }
\label{fig9}
\end{figure}

Before to make a comparison of our results with the data, let us analyse RPA
solutions in order to identify a wobbling mode.
Top right panels of Figs.\ref{fig8}, \ref{fig9} display four lowest RPA solutions of
Eq.(\ref{det}) and the redundant mode of Eq.(\ref{det}) as a function of the  
rotational frequency.
We recall that these solutions are found at different equilibrium deformations
(see Fig.\ref{fig2}). Indeed, in the both nuclei the criteria Eq.(\ref{crit})
uniquely determines the redundant and the wobbling modes. In Figs.\ref{fig8}, \ref{fig9} 
the redundant mode is manifested as a straight line (see the top right panels),
while the corresponding coefficient $c_\Omega$ is always -1 (the bottom panels).
The rotational mode is separated clearly from the vibrational modes.
Notice that the solutions, that are different from rotational and wobbling modes, 
contribute to the sum Eq.(\ref{sum}) with a zero weight, as it was proposed above.

To compare our results with available experimental data on
low lying excited states near the yrast line \cite{nndc}
we construct the Routhian function for each rotational band $\nu$
($\nu = yrast, \beta, \gamma, ...$)
\begin{equation}
R_{\nu}(\Omega)= E_{\nu}(\Omega) - \hbar \Omega I(\Omega), \quad \quad \Omega(I)=
\frac{E_{\nu}(I+1)-E_{\nu}(I-1)}{2}
\end{equation}
and define the experimental excitation energy in the rotating frame
$\hbar \omega_{\nu}^{\it exp} = R_{\nu}(\Omega) - R_{yr}(\Omega)$ \cite{n87}.
In $^{162}$Yb it is known only one negative signature $\gamma$-vibrational state.
The first RPA solution ($\nu=1$) is a negative signature gamma-vibrational
mode (with odd spins) till $\hbar\Omega\approx 0.28$MeV. With the increase of
the rotational frequency it is transformed to the wobbling
mode at $\hbar\Omega\approx 0.32$MeV (according to the criterion Eq.(\ref{crit})).
The other solutions ($\nu=2,3,4$) contribute to the sum Eq.(\ref{sum}) with a zero 
weight. Our results for $\nu=1$ solution may be used as a guideline for possible 
experiments on identification of the wobbling excitations near the yrast line. 
Although the positive signature states have been discussed in Ref.\onlinecite{JK05}, 
for completeness of our analysis we compare RPA results for the positive signature 
with updated database \cite{nndc}. According to our analysis, the first RPA 
solution of the positive signature may be identified with $\beta$-excitations at 
small rotation $\hbar \Omega\leq 0.2$MeV. With the increase of the rotational 
frequency a strong mixing between $\gamma$- (the second RPA solution at low spins) 
and $\beta$-excitations takes place. At $\hbar\Omega\geq 0.2$ Mev the first RPA 
solution of the positive signature   is determined by a single two-quasiparticle 
neutron configuration (see discussion and Table 1 in Ref.\onlinecite{JK05}).

In $^{156}$Dy the first ($\nu=1$) positive signature RPA solution 
carries a large portion of quasiparticle states with $K=2$.  
Since the quantum number K is reliable at small angular momenta, 
we associate the first positive signature  RPA solution with $\gamma$-vibrational mode. 
Rotation leads to a strong mixing between the first and the second RPA solutions. 
With the increase of the rotational frequency the first solutions  is separated, however, 
from the second one, while the latter one strongly interacts with the third RPA solution. 
At $\hbar \Omega\approx 0.3$MeV there is a crossing between B1 (yrast) band and B8 (excited) band 
which becomes the yrast after the transition point. The first positive signature RPA 
solution describes this transition with a good accuracy: the RPA mode goes to zero 
at $\hbar \Omega\approx 0.3$MeV. We recall that according to our analysis \cite{JK05}, 
namely this vibrational mode is responsible for the backbending phenomenon in this 
nuclei. 

The first negative signature RPA solution in $^{156}$Dy  can be associated with the negative 
signature gamma-vibrational mode with odd spins. After the transition from 
the axial to nonaxial rotation, at $\hbar\Omega\approx 0.3$MeV, according 
to the criteria Eq.(\ref{crit}), the first negative signature RPA solution describes 
the wobbling excitations.  The mode holds own features with the increase of 
the rotational frequency up to $\hbar\Omega \approx0.55$MeV. There is
a good agreement (see Fig.\ref{fig9}) between Routhians for this band and for 
the experimental band B10 (or $(+,1)_1$ band according to Ref. \onlinecite{Kon}). 
On this basis we propose to consider the B10 band as the wobbling band 
in the range of values $0.45\,MeV<\hbar \Omega < 0.55\,MeV$ 
($33\,\hbar \leq I \leq 39\,\hbar$ for this band). 
Note that the band B10 contains  the states with $31\,\hbar - 53\,\hbar$. 
However, our conclusion is reliable only for the states 
with $I=33\hbar-\,39\,\hbar$ (or up to $\hbar \Omega < 0.55\,MeV$). 
At $\hbar \Omega \approx 0.55\,MeV$ a crossing of the negative parity and 
negative signature (positive simplex) B6 band with the yrast band B8 is observed. 
Therefore, for $\hbar \Omega > 0.55\,MeV$ (or for $I>39\,\hbar$ for the B10 band)
one may expect an onset of octupole deformation in the yrast states.
Since  the octupole deformation is beyond the scope of our model, based on the
quadrupole deformed mean field, this feature will be discussed in forthcoming paper.

The proposed criterion Eq.(\ref{crit}) is a necessary but not
a sufficient condition to reach a conclusion that we have found a solution, related
to the wobbling excitations.
It is brought about by the formal equivalence between the classical Eq.(\ref{cwob})
and the microscopic Eq.(\ref{wob2}) equations for the wobbling mode.
Notice that our solution is determined in the UR frame, where the fluctuations of 
the angular momentum are absent (which are responsible for the wobbling mode in 
the PA frame).  
To identify the wobbling mode we have to specify also the relation between
electromagnetic transitions  in the Bohr-Mottelson model,
defined in the PA systems, and our model, defined in the UR frame.

\subsection{Electromagnetic transitions}
Transition probabilities for $X\lambda$ transition $|I\nu> \rightarrow |I'\nu'>$
between two high-spin states are given by expression
\begin{eqnarray}
\label{cleb}
&&B(X\lambda; I\nu \rightarrow I'\nu') \simeq \\
&&(I\,I\,\lambda\,\mu_{1}\,|\,I'\,I')^{2} |
<\nu'|\hat{\cal M}(X\lambda; \mu_{1}=I' - I)|\nu>|^{2}\nonumber
\end{eqnarray}
At high spin limit ($I >> \lambda$, $I' >> \lambda$), the transition 
from a one-phonon state into the yrast line state takes the form \cite{mar1}

\begin{eqnarray}
\label{be2}
&&B(X\lambda; I\nu \rightarrow I' yr))\simeq\\
&&|\langle RPA|\left[\hat{\cal M}^{(1)}(X\lambda; \mu_{1}=I' - I),
\hat O_{\nu}^\dagger \right]|RPA \rangle|^{2}\nonumber
\end{eqnarray}
Here, $\hat{\cal M}^{(1)}(X\lambda\mu_{1})$ is the linear boson part of the
corresponding transition operator of type $X$, multipolarity $\lambda$ and the
projection $\mu_{1}$ onto the rotation axis x in the UR frame.
The commutator in (\ref{be2}) can be easily expressed in terms of
phonon amplitudes $\psi^{(\nu)}_{\mu}$, $\phi^{(\nu)}_{\mu}$ (see Eq.(\ref{phon})).

With the aid of the transformation from x-to z-axis quantization \cite{mar1}
\begin{equation}
{\hat{\cal M}}(X\lambda \mu_{1}) =
\sum_{\mu_{3}} \,\,{\cal D}^\lambda_{\mu_{3} \mu_{1}}
(0,\frac{\pi}{2},
0) \,\,{\hat{\cal M}} (X \lambda \mu_{3}),
\label{mtf}
\end{equation}
and the definitions Eqs.(\ref{R12}),(\ref{x1}),
taking into account  that $\langle\nu|\hat M^{(E)}_{2\mu_{3}=0,2}|\nu\rangle =
\langle \hat M^{(E)}_{2\mu_{3}=0,2} \rangle$
holds in the first RPA order, we obtain
\begin{eqnarray}
&&B(E2; I\, \nu\, \rightarrow \, I \pm 1 yr) =
\left|\langle \left[\hat M_{2\mu_1=\pm1}^{(E)}, \hat O_\nu^\dagger\right]\rangle\right|^2=\nonumber\\
&&\left|\frac{i}{\sqrt{2}}\left[\tilde O_2^{(-)(E)},\hat O_\nu^\dagger \right]/\eta \mp
\frac{1}{\sqrt{2}}\left[\tilde O_1^{(-)(E)}, \hat O_\nu^\dagger \right]/\xi\right|^2=\nonumber\\
&&=\left|r_2^\nu B^{(E)}\mp r_1^\nu A^{(E)} \right|^2
\label{ntr}
\end{eqnarray}
Here, we use $\hat{M}^{(E)}=(eZ/A) \hat{M}$.
In virtue of Eq.(\ref{wob4}), we arrive to the expression for
the quadrupole transitions from the one-phonon wobbling state to the
yrast states
\begin{eqnarray}
&&B(E2; I\, w\, \rightarrow \, I \pm 1 yr) =
\frac{1}{4\langle {\hat J}_x \rangle}\times\nonumber\\
&&\times\Biggl| \left(\frac{W_2}{W_3}\right)^{\frac{1}{4}}A^{(E)}
\mp \left( \frac{W_3}{W_2} \right)^{\frac{1}{4}}B^{(E)}\Biggr| ^2
\label{b1}
\end{eqnarray}
which is similar to Eq.(\ref{rbe}). Thus, we provide a complete  microscopic definition
of the wobbling excitations in accordance with the
criteria suggested by Bohr and Mottelson for the rigid
rotor \cite{BM75}. Notice a clear difference between the microscopic and the
rigid rotor models: the microscopic moments of inertia, Eqs.(\ref{mmoi}),
should be calculated for the RPA solutions of Eq.(\ref{det})
that must obey to the condition Eq.(\ref{crit}).

For the intraband transitions we have (see Eq.(43) in Ref.\onlinecite{JK05})
\begin{eqnarray}
&&B(E2; I \nu \rightarrow I - 2 \, \nu) =
\left| \langle \nu|\hat{\cal M}(E2; \nu_{1}=2)|\nu\rangle \right|^{2} =\nonumber\\
&&= \frac{1}{8} \left|\sqrt{3}
\langle \hat{Q}^{(E)}_{0}\rangle -
\langle \hat{Q}^{(E)}_{2}\rangle \right|^{2}
\label{b2}
\end{eqnarray}

\begin{figure}[ht]
\includegraphics[height = 0.26\textheight]{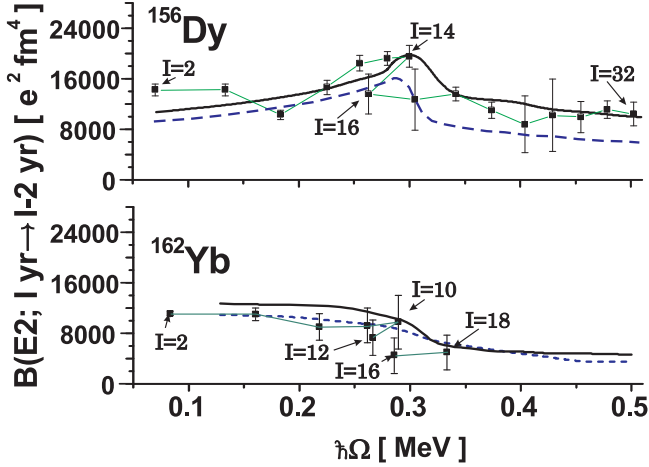}
\caption{
(Color online) Reduced transition probabilities 
$B(E2; I \, yr \, \rightarrow \, I-2 \, yr)$
along the yrast line. Experimental data (filled squares) 
are connected with a thin line to  guide
eyes. The results of calculations by means of Eq.(\ref{tbyr}) and of 
Eq.(\ref{b2}) are connected by dashed and solid lines, respectively.
}
\label{fig10}
\end{figure}
For illustrative purposes, to get a rough idea on the major trend of the 
quadrupole transitions, we employ the relations from the pairing-plus-quadrupole 
model (cf Ref.\onlinecite{fra})
\begin{eqnarray}
m\omega_0^2 \beta \,cos\gamma &=&
 \chi \langle Q_{0}\rangle\nonumber\\
m\omega_0^2 \beta \,sin\gamma &= -&\chi \langle Q_{2}\rangle
\label{Hartree}
\end{eqnarray}
between the deformation parameters $\beta$ and $\gamma$ and two intrinsic quadrupole
moments.
By dint of Eq.(\ref{Hartree})
and the definition of  the quadrupole isoscalar strength
$\chi= 4\pi m\omega^{2}_{0}/5\langle r^{2}\rangle \approx 4\pi m\omega^{2}_{0}/(3AR^{2})$
($R\approx1.2A^{\frac{1}{3}}fm$) one can express
the $E2$-transition probability via the deformation parameters $\beta$ and $\gamma$
\begin{eqnarray}
&&B(E2;\,I\,n_{w}=1 \rightarrow I\pm1 \, yr) = \Theta \times\\
&&\frac{\beta^2}{\langle \hat J_x\rangle}
\Biggl[\Biggl(\frac{W_2}{W_3}\Biggl)^{\frac{1}{4}} sin(\frac{\pi}3-\gamma) \pm
\Biggl(\frac{W_3}{W_2}\Biggr)^{\frac{1}{4}} sin \gamma \Biggr]^2 \nonumber\\
\end{eqnarray}
where $\Theta=(9/16\pi^{2})e^{2}Z^{2}R^{4}$.
Depending on the sign of $\gamma$-deformation
(for $W_{2,3} > 0$) one obtains  selection rules for
the quadrupole transitions from the one-phonon wobbling band to the yrast one
\begin{eqnarray}
a) &&For\, -60^{o}<\gamma<0 : \label{esr}\\
&&B(E2; I\,n_{w}\rightarrow I - 1 yr) >
B(E2; I\,n_{w}\rightarrow I + 1 yr)\nonumber\\
b) &&For\, 0<\gamma< 60^{o} :\nonumber\\
&&B(E2; I\,n_{w}\rightarrow I + 1 yr) >
B(E2; I\,n_{w}\rightarrow I - 1 yr)\nonumber
\end{eqnarray}

For the intraband transitions we obtain
\begin{equation}
B(E2;\,I\,n_{w}\rightarrow I-2 \, n_{w}) =\frac{1}{2}\Theta
\beta^2 cos^{2}(\frac{\pi}6-\gamma)
\label{tbyr}
\end{equation}
One observes from Eq.(\ref{tbyr}) that for the transitions along the yrast 
line ($n_w=0$) the onset of the positive (negative) values of $\gamma$-deformation 
leads to the increase (decrease) of the transition probability along the yrast line.

Experimental values of $B(E2, \, I \, \nu \, \rightarrow \, I^{\prime} \, yr)$
are deduced from the half life  of the yrast states \cite{nndc} using
the standard, long wave limit expressions
$B(E2, i \, \rightarrow \, f) =\
P(i\, \rightarrow \, f)/(1.223\times10^{9}E^{5}_{\gamma}) 
\,\, (e^{2}$ fm$^{4})$ \cite{RS}. Here, the transition energy 
$E_{\gamma}$ is in MeV and the absolute transition 
probability $P(i\, \rightarrow \, f)=\ln 2 /T(i\, \rightarrow \, f)$ is the
related to the half life $T(i\, \rightarrow \, f)$ (in seconds).
When comparing our results with experimental data, we take into account 
the Clebsh-Gordon coefficient (see Eq.(\ref{cleb})) up to $I\leq 10\hbar$.
For $I>10\hbar$ the asymptotic value for the Clebsh-Gordon coefficient,
which is 1, is used. Note that in the vicinity of the backbending the mean 
field description becomes less reliable (cf Ref.\onlinecite{ikuko}). While the 
cranking approach should be complemented with a projection technique in the 
backbending region due large fluctuations of the angular momentum (cf Ref.\onlinecite{RS}), 
its validity becomes much better at high spins. Evidently, the larger is the rotational 
frequency the stronger is a predictive power of the CRPA, since it is based on 
the cranking approach aimed for the high spin physics \cite{fra}.

Experimental data for the quadrupole transitions along the yrast line are compared 
with the results of calculations: (a) by means of Eq.(\ref{b2}) and calculations 
(b) by means  of Eq.(\ref{tbyr}) (see Fig.\ref{fig10}). In the calculations (a)  
we use the mean field values for the quadrupole operators. 
The calculations (a) evidently manifest the backbending effect obtained 
for the moments of inertia (see Figs.\ref{fig5},\ref{fig6})
at $\hbar \Omega_c \approx 0.25,\, 0.3$ MeV for $^{162}$Yb and $^{156}$Dy, 
respectively. Thus, the use of the self-consistent
expectation values $\langle \hat Q^{(E)}_{m}\rangle$ is crucial
to reproduce the experimental behaviour of the yrast band decay.
The calculations (b) (Eq.(\ref{tbyr})) reproduce the experimental data with
less accuracy, while providing the major trend of the transitions with the sign 
of $\gamma$-deformation. The agreement between calculated  and experimental values
of intraband $B(E2)$ transitions along the yrast line is especially good after 
the transition point.

\begin{figure}[ht]
\includegraphics[height=0.33\textheight,clip]{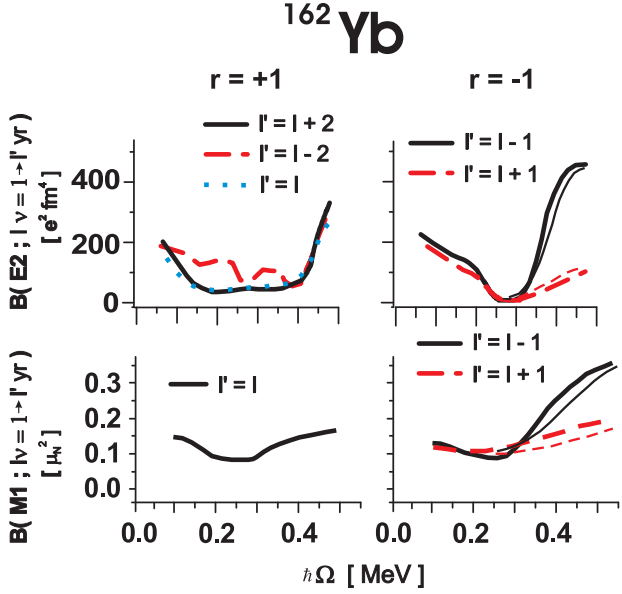}
\caption{(Color online)
$^{162}$Yb. The electric B(E2)- (top) and the magnetic B(M1)- (bottom) reduced 
transition probabilities from the one-phonon bands to the yrast band. The 
positive (negative) signature phonon band is described by 
the first $r=+1$ ($r=-1$) RPA solution.
One observes a strong dominance of the B(E2)- and B(M1)-transitions from
the wobbling states ($r=-1$) with spin I to the yrast states with spin
$I^\prime=I-1$, starting from the rotational frequency $\hbar \Omega > 0.28$MeV.
The transitions, calculated by means of the $\psi^{(\nu=1)}_{\mu}$ and 
$\phi^{(\nu=1)}_{\mu}$ phonon amplitudes,
are connected by solid lines. On the right panels the results obtained 
by means of Eqs.(\ref{b1}), (\ref{mmt}) (with the aid of the variables $W_{2,3}$,
Eq.(\ref{maw})) are connected by thin lines, starting from the rotational frequency 
$\hbar\Omega\approx 0.3$ MeV. This point is associated in our analysis with 
the appearance of wobbling excitations.
}
\label{fig11}
\end{figure}

\begin{figure}[ht]
\includegraphics[height=0.33\textheight,clip]{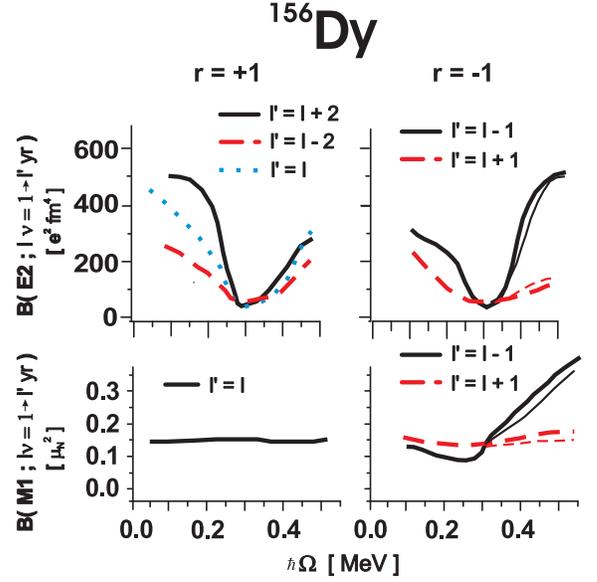}
\caption{(Color online)
$^{156}$Dy. Similar to Fig.\ref{fig11}.
One observes a strong dominance of the
B(E2)- and B(M1)-transitions from the wobbling states ($r=-1$) with spin I to
the yrast states with spin $I^\prime=I-1$ starting from the rotational
frequency $\hbar \Omega \geq 0.3$MeV.}
\label{fig12}
\end{figure}

At small rotational frequency, in the both nuclei, transitions probabilities 
from the first positive and negative signature RPA solutions are much weaker 
compare with the quadrupole transitions along the yrast line (see Fig.\ref{fig10}
and top panels of Figs.\ref{fig11},\ref{fig12}). At $\hbar\Omega \sim 0.05$MeV
the transition strengths for the first positive ($r=+1$) and negative ($r=-1$) 
RPA solutions are : $\sim 330 e^2 fm^4$ for $^{162}$Yb and $\sim 500 e^2 fm^4$ 
for $^{156}$Dy, with small differences between different transitions
due to the Clebsh-Gordon coefficients. On the other hand, we obtain a good 
correspondence between the shape evolution and the selection rules (\ref{esr}) 
for the both nuclei (see top panels of Figs.\ref{fig11},\ref{fig12} and 
Fig.\ref{fig2}). The transition probabilities, Eq.(\ref{ntr}), 
are calculated by means of the $\psi^{(\nu)}_\mu$ and $\phi^{(\nu)}_\mu$ phonon 
amplitudes expressed via the coordinate and momentum amplitudes 
(see Eqs.(\ref{phon}),(\ref{XP}),(\ref{R12}),(\ref{x1})). We compare 
these results for the first negative signature RPA solution (which is associated 
with a wobbling mode) with the results obtained with the aid of the Marshalek 
moments of inertia (see Eqs.(\ref{mmoi}),(\ref{maw}),(\ref{b1})). 
Evidently, if the "spurious" solution (the redundant mode) is not removed 
from Eq.(\ref{det}), it contributes to the variables (\ref{x1}). 
These variables can not obey to the condition (\ref{cond1}) in this case. 
As a result, the orthogonality condition Eq.(\ref{ort}) is broken and 
Eqs.(\ref{ntr}),(\ref{b1}) for the transitions 
should  produce completely different results. A good agreement between the both 
calculations (see the right top panels in Figs.\ref{fig11},\ref{fig12}) is the 
most valuable proof of the validity of our approach. The observed negligible 
differences are due the approximate fulfillment of the conservation laws 
(\ref{spur}), caused by the presence of the additional term (\ref{4a}) 
(see Fig.\ref{fig4}).

According to our analysis, at $\hbar \Omega \sim 0.25$MeV
there is a transition from the axially-deformed to nonaxial shapes
with the negative $\gamma$-deformation
in $^{162}$Yb (see Fig.\ref{fig2} and the discussion in Ref.\onlinecite{JK05}).
With a slight increase of the rotational frequency, starting from 
$\hbar \Omega \sim 0.28$MeV, the excited band of the negative signature,
created by the first RPA solution, changes the decay properties.
The negative values of $\gamma$-deformation
produce the dominance of the inter-band quadrupole transitions from
the one-phonon state  to the yrast ones with a lower spin ($\Delta I=1$, 
the case Eq.(\ref{esr}), a)).

Similar results are obtained in $^{156}$Dy for the lowest negative signature 
excited band, created by the first RPA solution. At low angular momenta 
($\hbar \Omega \leq 0.3$MeV) this band populates with approximately equal 
probabilities the yrast states with $I^\prime=I\pm1$ ($I$ is the angular momentum 
of the excited state). At $\hbar \Omega \sim 0.3$MeV a shape-phase transition 
occurs, that leads to the triaxial shapes with the negative $\gamma$-deformation. 
In turn, the excited band, created by the first RPA solution,
decays stronger on the yrast states with angular momenta $I^\prime=I-1$
($\Delta I=1$, the case Eq.(\ref{esr}), a)), starting from $\hbar\Omega\geq 0.32$MeV.

From the above analysis of the electric quadrupole transitions one
notices that there is no need to know the definition of the wobbling phonon
operator in the UR frame. Indeed,  in this frame the direction of the angular 
momentum is fixed and fluctuations of the angular momentum are absent.
However, there is a vibrational mode, related to the shape fluctuations, that carries
one unit of the angular momentum. According to Bohr and Mottelson consideration, 
in the PA frame the system shape is fixed, while the angular momentum fluctuates 
around the rotation axis that coincides with one of the principal axes of the 
inertia tensor. Evidently, the result for the transition probabilities
in the lab frame must be independent on the the choice of the reference frame.

To prove the equivalence of the both results for
the electric quadrupole transitions, let us use the Bohr-Mottelson definition of
the wobbling phonon operator, Eq.(\ref{wph}),
\begin{equation}
{\hat Q}_w^\dagger=\frac{i}{\sqrt{2\langle {\hat J}_x \rangle}}\Biggl[
(\hat I_2)_{PA}\Biggl(\frac{W_2}{W_3}\Biggl)^{\frac{1}{4}}+
(i\hat I_3)_{PA}\Biggl(\frac{W_3}{W_2}\Biggl)^{\frac{1}{4}}\Biggr].
\label{wbm}
\end{equation}
Here, the quantities $W_{2,3}$
are determined by the Marshalek moments of inertia, Eqs.(\ref{maw}).
In the PA frame  one has to use the transformation Eq.(\ref{bdf}) in order
to calculate the transition probability and commutation relations Eq.(\ref{rcom}).
This transformation and the definition (\ref{wbm})
yield the expression
\begin{eqnarray}
&&B(E2; I\, w\, \rightarrow \, I \pm 1 yr) =
\left|\langle \left[\hat M_{2\mu_1=\pm1}^{(E)}, \hat O_w^\dagger\right]\rangle\right|^2=\nonumber\\
&&
\frac{1}{4\langle {\hat J}_x \rangle}
\Biggl| \left(\frac{W_2}{W_3}\right)^{\frac{1}{4}}A^{(E)}
\mp \left( \frac{W_3}{W_2} \right)^{\frac{1}{4}}B^{(E)}\Biggr| ^2
\end{eqnarray}
that is similar to Eq.(\ref{b1}), obtained in the UR frame, indeed. 
We will use this fact below to understand major features of the magnetic transitions
from the wobbling band.

In the CRPA approach the magnetic transitions  are defined as
\begin{equation}
B(M1; I\,\nu \rightarrow \, I\pm 1 \, yr) \approx
\left|\langle \left[\hat M_{1\mu_1=\pm1}^{(M)}, \hat O_\nu^\dagger\right]\rangle\right|^2
\label{mag}
\end{equation}
With the aid of the transformation from x-to z-axis quantization Eq.(\ref{mtf}) 
one obtains
\begin{equation}
B(M1; I\, \nu \rightarrow \, I\pm 1 \, yr) \approx
\frac{1}{2}\left|i \left[{\hat M}^{(M)}_{1\nu_{3}=1},{\hat O}_\nu^\dagger\right]
\mp \left[{\hat M}^{(M)}_{1\nu_{3}=0},{\hat O}_\nu^\dagger\right]\right|^2
\end{equation}
The linear boson term  of the magnetic operator has 
a form (see also Ref.\onlinecite{JK})
\begin{eqnarray}
&&\hat M^{(M)}_{1\nu_{3}=0,1} = \mu_{N} \sqrt{3} \left( \frac{1}{2} g_{s}^{(eff)} 
\hat S_{01\nu_{3}} + g_{l}^{(eff)} \hat L_{01\nu_{3}}\right)\nonumber\\
&&=\frac{i^{\nu_{3}+2}}{2} \sum_{\mu} 
\Lambda^{(\nu_{3})}_{\mu}(b^{+}_{\mu} + (-1)^{\nu_{3}} b_{\mu}),
\end{eqnarray} 
where $\mu_{N}$ is the nucleon magnetons, $g_s^{(eff)}$, $g_l^{(eff)} $
are the spin and orbital effective gyromagnetic ratios, respectively; and 
the quasiparticle matrix elements $\Lambda^{(\nu_{3})}_{\mu}$ are real.
Taking into account the definition of the phonon operator Eq.(\ref{phon}), we
express the magnetic transition with the aid of generalized coordinate and 
momentum amplitudes Eqs.(\ref{XP})
\begin{equation}
B(M1; I\, \nu \rightarrow \, I\pm 1 \, yr) \approx
\left|\sum_\mu \Lambda^{(1)}_{\mu}X_\mu^\nu 
\mp \sum_\mu \Lambda^{(0)}_{\mu}P_\mu^\nu \right|^2
\end{equation}

Are there any specific features of magnetic transitions from the wobbling band, 
related to the sign of $\gamma$-deformation ?  Our results evidently demonstrate 
the dominance of $B(M1; I\,n_{W}\rightarrow I - 1 yr)$ (see the right bottom 
panels in Figs.\ref{fig11},\ref{fig12}) for the both nuclei. To get an insight 
of this result, we define the magnetic transitions by dint of the Marshalek moments 
of inertia. With aid of the definition, Eq.(\ref{wbm}), the transformation, 
Eq.(\ref{bdf}), and commutation relations 
$[\hat I_1\pm i\hat I_2, \hat M_{\lambda m}]=\sqrt{\lambda(\lambda+1)-m(m \mp 1)}\hat M_{\lambda m \mp 1}$
for the PA frame \cite{BM1}, we obtain from Eq.(\ref{mag})
\begin{eqnarray}
&&B(M1; I\, \nu\, \rightarrow \, I\pm 1 \, yr) \approx
\biggl|\langle \hat M^{(M)}_{1\nu_{3}=1}[r=+1]\rangle\biggr|^{2}\times\nonumber\\
&&\times
\frac{1}{4\langle {\hat J}_x \rangle}
\frac{(\sqrt{W_{3}}\mp\sqrt{W_{2}})^{2}}{\sqrt{W_{2}W_{3}}}
\label{mmt}
\end{eqnarray}
Although the expression for the magnetic transitions, Eq.(\ref{mmt}), is
similar to the one of the Bohr-Mottelson model, we stress that the moments of
inertia are defined self-consistently within the CRPA approach.
Notice that the dipole magnetic moment $\langle \hat M^{(M)}_{1\nu_{3}=1}[+] \rangle$
increases quite drastically, if a nucleus is undergoing the backbending 
(see the discussion about $M1$ strength in Ref.\onlinecite{JK2}).
Keeping in mind that for the wobbling states $W_{2,3}>0$, we have
\begin{equation}
B(M1; I\,n_{W}\rightarrow I - 1 yr) > B(M1; I\,n_{W} \rightarrow I + 1 yr)
\label{msr}
\end{equation}
Thus, the tendency observed in the microscopic calculations with the aid of 
the phonon amplitudes are understood in terms of the rules Eq.(\ref{msr}).
Independently on the sign of the $\gamma$-deformation 
of rotating nonaxial nuclei, these rules  determine the dominance of $\Delta I=1 \hbar$ 
magnetic transitions from the wobbling to the yrast states.

\section{Summary}

The observation of the wobbling excitations among excited states near the yrast line
is one of the most convincing manifestations of the nonaxial deformation in rotating
nuclei. In this paper we present a transparent, self-consistent derivation of
the basic equations for the wobbling excitations in the UR (time-independent) frame,
that determine: i) the energy spectrum and ii)electromagnetic properties of these
states in even-even nuclei. 
We obtained the same  expressions for the effective moments of 
inertia (\ref{mmoi}) as those obtained by Marshalek in the 
time-dependent Hartree-Bogoliubov approach in the PA frame \cite{mar2}.
We established one-to-one correspondence between the main characteristics of
the wobbling excitations in the Bohr-Mottelson model and those that are derived
within  the CRPA approach. 
Note, however, that the CRPA breaks down at the transition point 
when $\Delta_p$ or $\Delta_n$ vanish \cite{mar1}. We have avoided this problem by 
means of the phenomenological prescription for the rotational dependence of the 
pairing gap. A good agreement between the dynamic moment of inertia calculated 
at the mean field approximation and the Thouless-Valatin moment of inertia calculated
in the RPA supports the consistency of our mean field calculations (see Fig.\ref{fig3}).
In contrast to the standard RPA calculations,   
where the residual strength constants are fixed for all values of $\Omega$ 
(see e.g. \cite{shi1,mat1,mat2}), we determine the strength constants
for each value of $\Omega$ by the requirement of the validity of the 
conservation laws. This enables us to overcome the instability 
of the RPA calculations at the transition region, at least, for the excitations.
In principle, projection methods may be used in the transition region 
in order to calculate transition matrix elements. 
Although the amplitudes $\phi^{(\nu)}_{\mu}$ (see Eq.(\ref{phon})) 
are higher for the RPA modes in the transition
region than in other regions, the relation  
$|\phi^{(\nu)}_{\mu}|<|\psi^{(\nu)}_{\mu}|$ is still valid
to hold the QBA. And least but not last,
the CRPA becomes quite effective at high spins, after the transition point, when 
the pairing correlations still persist.  

From our analysis it follows that  an excited band can be considered as
the wobbling one, if the magnetic transitions from this band into the yrast one
fulfil the condition Eq.(\ref{msr}). Note that these rules do not depend on the
sign of $\gamma$-deformation.
In contrast, the electric quadrupole transitions 
from this band to the yrast one must fulfil the staggering rules Eq.(\ref{esr}) 
depending on the sign of  $\gamma$-deformation. The transition from this band are 
characterised by a high collectivity.
We predict that the lowest excited  negative signature and positive
parity band in $^{162}$Yb (which is a natural prolongation of the odd angular
momentum part of the $\gamma$-band) transforms to the wobbling band at
$\hbar\Omega>0.3$MeV. We found that  strong  E2 transitions
from this band  populate the yrast states, with the
branching ratio $B(E2; I\, w\, \rightarrow \, I - 1 yr)/B(E2; I\, w\, \rightarrow \, I +1 yr)>1$.
Such a behaviour is brought about by the onset of nonaxial nuclear shapes 
with the negative sign of $\gamma$-deformation after the backbending. 
According to our definition of $\gamma$-deformation, with the increase of the rotational 
frequency the system is driving to a noncollective oblate rotation (around the x-axes).
This trend is confirmed by a good agreement between the results for the quadrupole
transitions along the yrast line and available experimantal data (see Fig.\ref{fig10}).   
Similar transition occurs in $^{156}$Dy  after the backbending as well, at 
$\hbar\Omega>0.3$MeV. In  this nucleus we found that the lowest
negative signature and positive parity band represents a natural prolongation of
the $\gamma$-band with odd spins, observed at small angular momenta up to 
$\hbar\Omega\sim 0.3$MeV. At rotational frequency $\hbar\Omega> 0.3$MeV this
band created by the first RPA solution transforms to the wobbling excitations.
A good agreement between calculated in the CRPA  and experimental 
Routhians allows us to conclude that the experimental states, associated with 
$(+,1)_1$ band in Ref.\onlinecite{Kon}, are wobbling excitations at the rotational
frequency values $0.45$MeV$<\hbar\Omega<0.55$MeV. 
These states fulfil all requirements that are specific for the wobbling excitations  of 
nonaxially deformed rotating nuclei, with the negative $\gamma$-deformation.
Evidently, it is quite desirable to obtain new experimental data on  
electromagnetic decay properties of these states in order to reach 
a final conclusion about our prediction and, respectively, on the validity of
the CRPA analysis.

\section*{Acknowledgments}
This work is a part of the research plan MSM
0021620834 supported by the Ministry of Education of the Czech Republic
and by the project 202/06/0363 of Czech Grant Agency.
It is also partly supported by Grant No. FIS2005-02796 (MEC, Spain).
R. G. N. gratefully acknowledges support from the
Ram\'on y Cajal programme (Spain).

\appendix
\def\theequation{\thesection.\arabic{equation}}
\section{Asymmetric rotor model}
\setcounter{equation}{0}

In this appendix we review the basic features of the wobbling excitations
in the rotor model \cite{BM75} in order to compare with
the microscopic model discussed in section III. In addition to well known results,
we provide a novel analysis of magnetic properties of the wobbling states.

At high-spin limit $I>>1$, the rotor Hamiltonian has the following form
(see Ref.\onlinecite{BM75})
\begin{eqnarray}
&\hat{H} = \hat{H}_{0} + \frac{\hbar^2}{2 \Im_1} (\hat{I})^2_{PA} + \hat{H}_{wobb}
\label{hrot}\\
&[\hat I_i, \hat I_j]_{PA}=-i\varepsilon_{ijk}(\hat I_k)_{PA}
\label{rcom}
\end{eqnarray}
where $(\hat I_i)^{2}_{PA}$ and $\Im_i$ are the angular momentum and principal
moment of inertia components in the rotating, principal axis (PA) coordinate
system. It is assumed that the yrast band
is generated by rotation around the x-axis $I\approx I_1=K$. Small oscillations of
the angular momentum create wobbling excitations described by the term
\begin{eqnarray}
\hat{H}_{wobbl}& = &\hbar \, \omega_{w} \,\,\left( \hat Q_{wobb}^+ \hat Q_{wobb} \,+
\,\frac{1}{2} \right)
\nonumber\\
\hbar \omega_{w}& = &\hbar^2 I \sqrt{W_2 W_3}
\label{cwob}
\end{eqnarray}
where
\begin{equation}
W_2 = \frac{1}{\Im_2} \,-\,\frac{1}{\Im_1}\,, \qquad
W_3 = \frac{1}{\Im_3} \,-\,\frac{1}{\Im_1}
\end{equation}
The wobbling (excited) state at spin $I$ is created by the wobbling phonon
\begin{equation}
\label{wph}
\hat O^{+}_{wobb} = x B^+ - y B
\, \Leftrightarrow \,
 B^+ = x \hat O^+_{wobbl} + y \hat O_{wobbl}
\end{equation}
with
\begin{eqnarray}
 B^+&=&\frac{i}{\sqrt{2I}}(\hat I_2+i\hat I_3)_{PA},\quad B=( B^+)^+,\\
x&=& \frac{\sqrt{W_3} +\sqrt{W_2}}{2} (W_2 W_3)^{-1/4} \label{x}\\
y &=& \frac{\sqrt{W_2} -\sqrt{W_3}}{2} (W_2 W_3)^{-1/4}
\label{y}
\end{eqnarray}
and a normalization condition $x^2-y^2=1$.
From Eq.(\ref{cwob}) it follows that the diagonalization of $\hat H_{wobb}$ requires
$W_2 > 0$, $W_3 > 0$ and $\Im_1>\Im_2>\Im_3$.

At $(I \approx I_1 >> 0)$ the eigenfunction of Hamiltonian (\ref{hrot}) is a product
of the Wigner $\cal D$~-~function and the intrinsic eigenfunction defined by 
the wobbling quantum number $|n_{w}\rangle= (Q^{+}_{wobb})^{n_{w}}~|0\rangle $.
The variable $n_w$
\begin{equation}
n_w = \langle IK| B^+B|IK\rangle \approx I-I_1=I-K
\label{nw}
\end{equation}
is defined with respect to the state $|IK=(I_1)_{PA}\rangle\equiv|n_w\rangle$
such that $B^+|IK\rangle=\sqrt{n_w+1}|IK-1\rangle=\sqrt{n_w+1}|n_w+1\rangle$ and
$|0\rangle=|IK=I\rangle$.
The transition probability for the operator of type $X$ and multipolarity
$\lambda$
\begin{eqnarray}
&&B(X\lambda;I \nu K_{\nu} \rightarrow I'\nu'K_{\nu}')= \nonumber\\
&&\frac{1}{2I+1}\,\,|\langle I'K'_{\nu}\nu'||\hat{\cal M}(X\lambda)||IK_{\nu}\nu\rangle|^{2}
\end{eqnarray}
is defined by the reduced matrix element
\begin{eqnarray}
&&\langle I'K'n_{w}'||\hat{\cal M}(X\lambda)||IKn_{w}\rangle=\sqrt{2I+1}\times
\label{mel}\\
&&\sum^{\lambda}_{\nu_{1}=-\lambda}
\langle n_{w}'|(IK\lambda\nu_{1}|I'K')\hat{\cal M}(X\lambda\nu_{1}=I'-I)|n_{w}\rangle
\nonumber
\end{eqnarray}
Notice that the eigenmodes of the Hamiltonian $\hat H_{0}$
do not change the projection $K$ onto
the first PA axis. However, if the wobbling mode is excited
with $|n_{w}'-n_{w}| \neq 0$, the projection $K$ is changed and
corresponding Clebsch-Gordan coefficients in Eq.(\ref{mel}) can be expressed
in terms of $B^+$, $B$ operators
(see Section 4.5c in Ref.\onlinecite{BM75}).

Using this procedure for the  Clebsch-Gordan coefficients,
a relation between multipole operators in
the PA frame with the x- and z-quantization axis and the definition Eq.(\ref{xz})
\begin{equation}
{\hat {\cal M}}(X\lambda \mu_{1}) =
\sum_{\mu_{3}} \,\,{\cal D}^\lambda_{\mu_{1} \mu_{3}}
(0,\frac{\pi}{2},
0) \,\,{\hat {\cal M}} (X \lambda \mu_{3}),
\label{bdf}
\end{equation}
we obtain for interband E2 transitions between the one-phonon
wobbling band ($n_{w} = 1$) and the yrast band ($n_{w} = 0$)
\begin{equation}
B(E2;\,I\,n_{w}\rightarrow I\pm1 \,yr) =|\varrho_1 A^{(E)} \mp
\varrho_2 B^{(E)}|^2
\label{rbe}
\end{equation}
Here, the variables $A^{(E)}$ and $ B^{(E)}$ are defined by Eqs.(\ref{sym1}),
(\ref{sym2}), $\hat Q^{(E)}_m\equiv e\frac{Z}{A} \hat Q_m$
and
\begin{equation}
\varrho_1=\frac{1}{\sqrt{2I}}\left(\frac{W_2}{W_3}\right)^{\frac{1}{4}}, \quad
\varrho_2=\frac{1}{\sqrt{2I}}\left(\frac{W_3}{W_2}\right)^{\frac{1}{4}}
\label{clr}
\end{equation}
For intraband transitions we have

\begin{equation}
B(E2;\,I\,n_{w} \rightarrow I-2 \,n_{w}) =
\frac{1}{8}\left| \sqrt{3} \langle \hat Q^{(E)}_{0}\rangle
-\langle \hat Q^{(E)}_{2}\rangle\right|^2
\label{clr1}
\end{equation}

Here, we use the fact that $\langle\hat{\cal M}(E2,\nu_{1}=\pm1)\rangle = 0$ and
$\langle \hat{\cal M}(E2,\nu_{1}=2)\rangle = \langle\hat{\cal M}(E2,\nu_{1}=-2)\rangle$
and neglect terms of order or higher than $\sim \frac{1}{I}$.

Similar procedure, described above, can be employed to derive M1 transitions from 
one-phonon wobbling band into the yrast band. Ignoring the terms of order 
$I^{-1}$ in Eq.(\ref{mel}) at high-spin limit ($I\approx K >> 1$), we have the 
following approximative values for the Clebsch-Gordan coefficients in terms of the 
matrix elements of the operators $B$, $B^{+}$ (or $\hat Q_{wobb}$, $\hat Q^{+}_{wobb}$)
\begin{eqnarray}
\nonumber
(I\,K\,1\,0\,|\,I\,K) &\rightarrow&  \,\,1  \\
\nonumber
(I\,K\,1\,0\,|\,I\pm1\,K) & \rightarrow & \pm \frac{1}{\sqrt{I}} \langle n \pm 1|
\begin{array} {c}
B^{+}\\ B
\end{array}|n\rangle \\
\nonumber
(I\,K\,1\,\pm1\,|\,I\,K\pm1) & \rightarrow & \mp \frac{1}{\sqrt{I}}
\langle n \mp 1|
\begin{array} {c}
B\\ B^{+}
\end{array}
|n\rangle \\
(I\,K\,1\,\pm1\,|\,I\pm1\,K\pm1) & \rightarrow & 1
\end{eqnarray}
Since $\langle \hat {\cal M}(M1 \,\nu_1=\pm 1)\rangle=0$,
we obtain
\begin{eqnarray}
&&\langle I'K'n_{w}'||\hat {\cal M}(M1 \,\nu_1=0)||IKn_{w}\rangle =\nonumber\\
&& \sqrt{2I+1}
\langle n'_{w}|\Biggl[ \delta_{I',I} \langle \hat{\cal M}(M1,\nu_1=0) \rangle + \\
&+&\,\delta_{I',I+1} \frac{1}{\sqrt{I}} \langle \hat{\cal M}(M1,\nu_1=0)\rangle\,
(x\hat Q^{+}_{wobb}+y \hat Q_{wobb}) \nonumber\\
&-& \delta_{I',I-1} \frac{1}{\sqrt{I}} \langle \hat{\cal M}(M1,\nu_1=0)\rangle\,
(x\hat Q_{wobb}+y\hat Q^{+}_{wobb}) \Biggr]|n_{w}\rangle\nonumber
\end{eqnarray}

Here,  ${\hat{ \cal M}}(M1;\nu_{1}) =
\mu_N\,\sqrt{3} \sum_{i=1}^{A}$  $($  $\frac{1}{2} g_s^{(i,eff)} [\sigma
\otimes Y_{l=0}]_{1\nu_{1}} \,+\,$
$g_l^{(i,eff)} [l \otimes Y_{l=0}]_{1\nu_{1}}$ $)$ is a magnetic dipole operator.
With the aid of Eqs.(\ref{x}), (\ref{y}),(\ref{bdf}) and the definition of operators
given in Ref.\onlinecite{JK}, we obtain for M1 transitions from the one-phonon
wobbling band ($n_{w}=1$) into the yrast band ($n_{w}=0$) the following expression
\begin{eqnarray}
B(&M1&; I\,n_{w}\rightarrow I\pm1 yr) =
\biggl|\langle \hat M^{(M)}_{1\nu_{3}=1}[r=+1]\rangle\biggr|^{2}\times\nonumber\\
&&\times \frac{1}{4I}\frac{(\sqrt{W_{3}}\mp\sqrt{W_{2}})^{2}}{\sqrt{W_{2}W_{3}}}
\label{mtr}
\end{eqnarray}
The magnetic moment
$\langle \hat M^{(M)}_{1\nu_{3}=1}[+]\rangle$ can be calculated in any
microscopic model.

\end{document}